\documentclass[pdflatex,sn-mathphys-num]{sn-jnl}
\usepackage{dcolumn}
\usepackage{bm}
\usepackage{graphicx}%
\usepackage{multirow}%
\usepackage{multicol}%
\usepackage{amsmath,amssymb,amsfonts}%
\usepackage{amsthm}%
\usepackage{mathrsfs}%
\usepackage[title]{appendix}%
\usepackage{xcolor}%
\usepackage{textcomp}%
\usepackage{manyfoot}%
\usepackage{booktabs}%
\usepackage{algorithm}%
\usepackage{algorithmicx}%
\usepackage{algpseudocode}%
\usepackage{listings}%
\usepackage{soul} 

\theoremstyle{thmstyleone}%

\theoremstyle{thmstyletwo}%

\theoremstyle{thmstylethree}%

\begin{document}

\title[Article Title]{XpookyNet: Advancement in Quantum System Analysis through Convolutional Neural Networks for Detection of Entanglement}

\author*[1]{\fnm{Ali} \sur{Kookani}}\email{ali.kookani@ut.ac.ir}
\author[2]{\fnm{Yousef} \sur{Mafi}}\email{yousef.mafi@ut.ac.ir}
\author[2]{\fnm{Payman} \sur{Kazemikhah}}
\author[3]{\fnm{Hossein} \sur{Aghababa}}
\author[1]{\fnm{Kazim} \sur{Fouladi}}
\author[4]{\fnm{Masou\emph{}d} \sur{Barati}}

\affil[1]{\orgdiv{Faculty of Engineering}, \orgname{College of Farabi, University of Tehran}, \orgaddress{\state{Tehran}, \country{Iran}}}

\affil[2]{\orgdiv{School of Electrical and Computer Engineering}, \orgname{College of Engineering, University of Tehran}, \orgaddress{\state{Tehran}, \country{Iran}}}

\affil[3]{\orgdiv{Department of Engineering}, \orgname{Loyola University Maryland}, \orgaddress{\state{Maryland}, \country{USA}}}

\affil[4]{\orgdiv{ Swanson School of Engineering, Electrical Engineering}, \orgname{University of Pittsburgh}, \orgaddress{\city{Pittsburgh} \state{Pennsylvania}, \country{USA}}}

\abstract{Quantum system attributes, notably entanglement, are indispensable in manipulating quantum information tasks. Ergo, machine learning applications to help harness the sophistication of quantum information theory have surged. However, relying on unsuited prototypes and not bridling quantum information data for usual processors has often resulted in sub-optimal efficaciousness and inconvenience. We develop a custom deep convolutional neural network, XpookyNet, which is streamlined with respect to the interrelationships of density matrices of two-qubit systems to underpin three-qubit systems, breaking new ground to more qubit systems by their subsystems. Comparative implementation of XpookyNet provides instantaneous and meticulous results with an accuracy of 98.53\% in merely a few epochs. XpookyNet effectively handles the inherent complexity of quantum information, equipping deeper insights into many-body systems as a bedrock. The study also investigates quantum features and their relation to the purity of a density matrix directly related to noise and system decoherence in NISQ-era quantum computation. Preparing the density matrix in a compact and compatible format with customary convolutional neural networks plays a determining role in dissecting quantum features. The procedure renders a convex criterion that detects entanglement and is a yardstick for quantum system coherence.}


\maketitle
\section{Introduction}\label{sec1}
In quantum mechanics, an uncanny phenomenon known as quantum entanglement arises when two or more particles interact so that their quantum states become related in a way that the neither state $|\psi_{AB}\rangle \neq |\psi_A\rangle \otimes |\psi_B\rangle$ nor $\langle A\otimes B\rangle\neq\langle A\rangle\otimes\langle B\rangle$ \cite{laloë2019, blasiak2019}. Instantaneous changes in one particle are simultaneously mirrored in others, irrespective of their spatial separation, with the non-locality condition of $P(A=a, B=b) \neq P(A=a) \cdot P(B=b)$ holding \cite{das2022}. Quantum computing needs entanglement to form and rise, as it allows the system to perform multiple calculations simultaneously using many qubits, achieves secure and high key rates, and underpins quantum annealing and variational quantum algorithms to solve intractable problems and optimization problems \cite{mooney2019, srinivas2021highfidelity, Yin2020Entangl, morstyn2022annealing, diezvalle2021quantum}.

The first step in manipulating systems is to detect the presence and amount of entanglement, so various entanglement detection criteria have been proposed \cite{Liu2022Detecting, PhysRevA.101.052117, Zangi2021Combo}. However, the positive partial transpose (PPT) criterion can determine entanglement with certainty for systems $\mathcal{H}_A$ and $\mathcal{H}_B$ where the Hilbert space is $\mathcal{H}_{AB} = \mathcal{H}_A \otimes \mathcal{H}_B$ with the necessary size of $2\otimes2$ or $2\otimes3$ non-mixed states. In other words, Hilbert space includes entangled mixed states that still meet the PPT conditions called bound-entangled states \cite{Huber2018High}.

Comprehensively assessing quantum entanglement requires choosing the right criterion. While criteria  like concurrence, negativity, and relative entropy of entanglement have advantages, the entanglement of formation (EoF) stands out for its operational interpretation of the resources needed to create a specific state. Unlike its counterparts, it provides a deeper understanding of entanglement phenomenon \cite{Bai2014General, Kim2021Entanglement, Wang2022}.

\begin{figure}[t]
\centering
\includegraphics[width=1\textwidth]{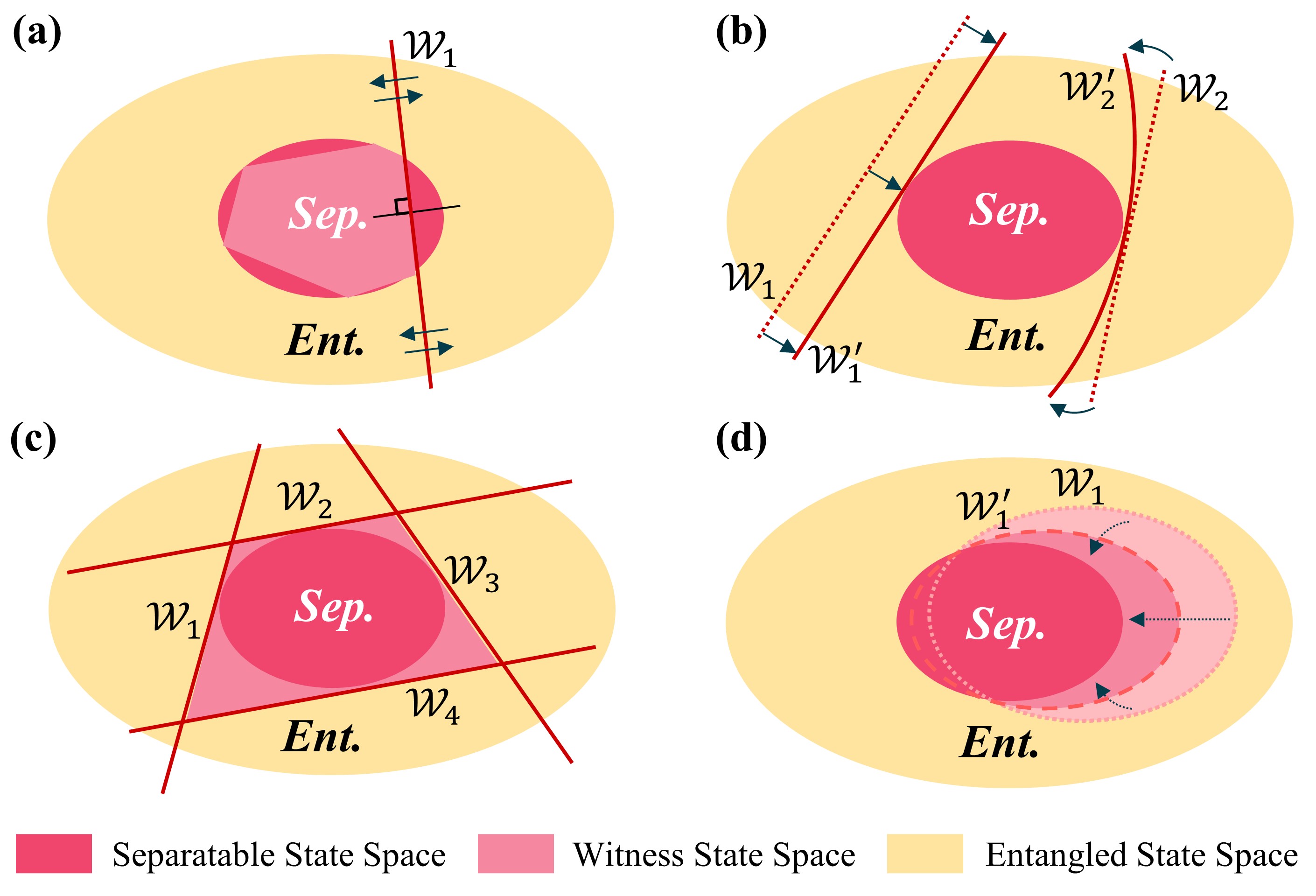}
\caption{Entangled or separable in Hilbert space. (a) Adjustment of a witness as a linear hyperplane and its optimization failure. (b) Entanglement witness optimization approaches, including linear: from $\mathcal{W}_1$ to $\mathcal{W}_1'$, and nonlinear: from $\mathcal{W}_2$ to $\mathcal{W}_2'$. (c) Hindrances to precision due to the convexity of the target space. (d) Improvement of naive ML usage as a convex witness isolator resulting in an ungeneralized model.}\label{Fig.1}
\end{figure}

Quantifying and classifying entanglement beyond three qubits is an NP-Hard problem. The generalized concurrence still measures entanglement, but its calculation is computationally intensive for mixed states and it cannot pinpoint individual qubits \cite{Bhaskara2016Generalized}. Entanglement witnesses are another practical tool to detect entanglement in many-body quantum systems. A witness $\mathcal{W}$ is a Hermitian operator that for all separable states, $\langle \mathcal{W}_{\rho_{sep}} \rangle = \text{Tr}(\mathcal{W} \rho_{sep}) \geq 0$ but for some entangled states $\langle \mathcal{W}_{\rho_{ent}} \rangle = \text{Tr}(\mathcal{W} \rho_{ent}) < 0$ \cite{arkhipov2018}. It detects entanglement without fully characterizing the system or performing a tomography; however, due to the exponential growth of variables with higher qubit counts, it requires optimization in high-dimensional spaces. Quantum witnesses can be optimized using machine learning (ML) because they can quickly identify patterns in large datasets, making them ideal for solving complex problems \cite{ma2018,lu2018}. Illustrations in Fig. \ref{Fig.1} give a grasp of compartmentalizing methods for classifying states \cite{hyllus2006,qi2012}.

Convolutional neural networks (CNNs) have revolutionized computer vision. They exploit hierarchical layers of convolution operations to exploit structured patterns in 3D tensors like images. To detect more complex structures, CNNs use convolutional layers as filters to identify edges, textures, and other details. By using hierarchical approaches, CNNs are able to understand spatial hierarchies and correlations within structured data \cite{Li2020A}.

Density matrices of quantum states share a structured nature analogous to the grid structure of images. They epitomize the coherence of different quantum states, similar to pixels in an image, arranged in a grid-like fashion. The application of CNNs to density matrices enables the identification of critical quantum properties and uncovers patterns that might be difficult to discern with traditional analyses \cite{Ahmed2020ClassificationAR}.

The link between ML and quantum information (QI) has recently been extensively studied \cite{qiu2019, harney2021, girardin2022}. There has yet to be a model with sufficiently rigorous accuracy for the two-qubit entanglement detection problem, and the design of CNN models for QI applications remains relatively unexplored. The research must characterize quantum system states more comprehensively beyond two qubits under the entanglement categories. 

\begin{figure}[t]
\centering
\includegraphics[width=1\textwidth]{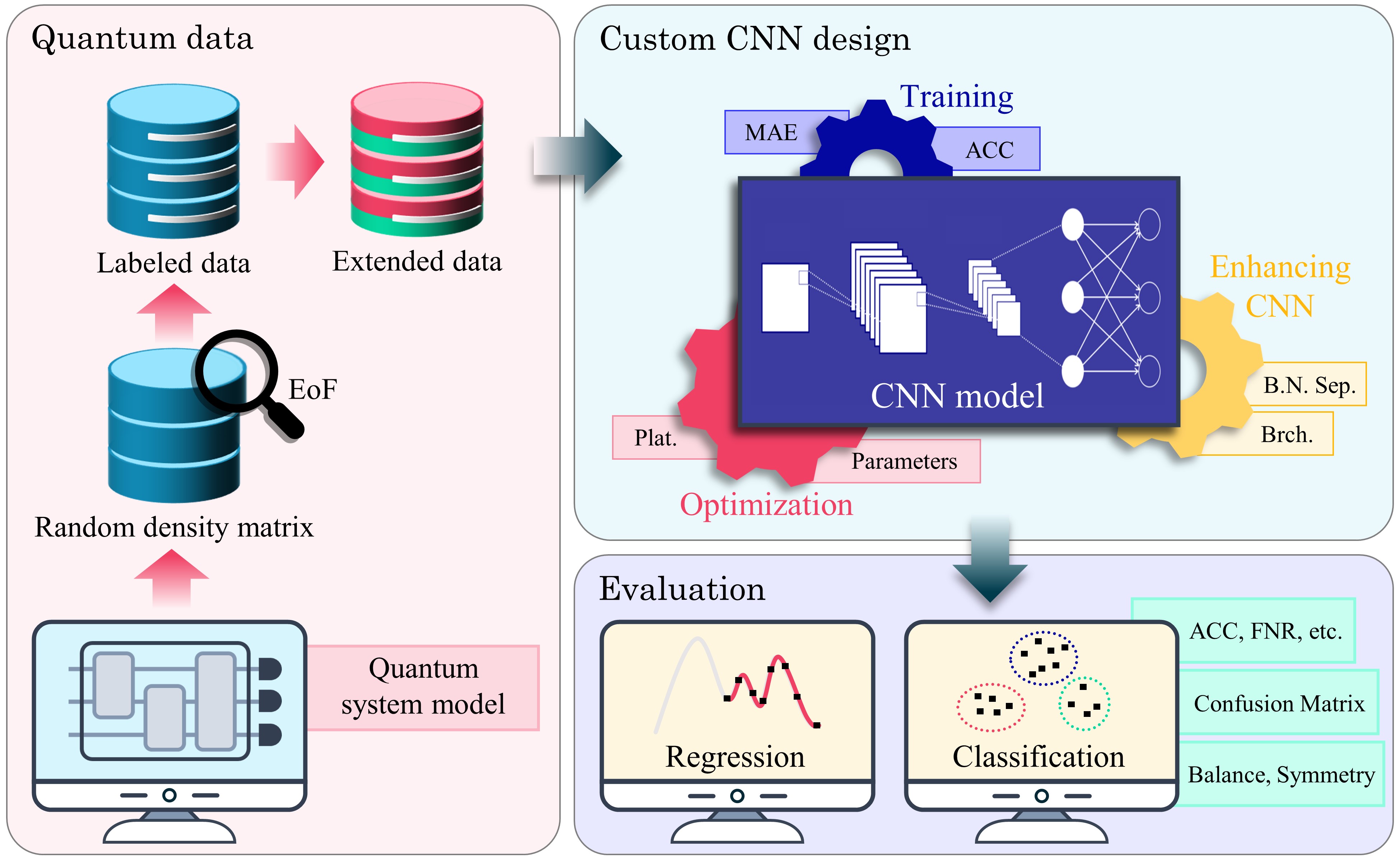}
\caption{A comprehensive overview of the study's scheme.}\label{Fig.2}
\end{figure}

Figure \ref{Fig.2} shows the research outline of this study. Initially, we create random density matrices and use them for training CNN. The density matrices produced for the data set are  labeled  using entanglement of formation (EoF). The EoF underlies the classification of the matrices on the basis of their characteristics. The matrices are then pre-processed into an extended-tensor format to prepare them for subsequent processing, as explained in section \ref{Sec.2}. In section \ref{Sec.3}, we develop custom CNN models. The CNN model is designed and refined through a training and optimization process. Plateau machination, data normalization, and branching features are added to enhance the model structure. The model is then subjected to testing to evaluate its performance. The resulting classification outcomes are used to assess the model's applicability and make necessary adjustments. This approach ensures an accurately designed, trained, optimized, and tested CNN model. Section \ref{Sec.4} involves constructing a confusion matrix to assess the balance and symmetry of the results. It helps to understand the accuracy of the model and the errors made. Key metrics like ACC, MCC, and FNR are calculated to analyze the study's performance and reliability extensively.

\section{Quantum Data}\label{Sec.2}
There are two methods to combine quantum and classical computers for ML. The first method involves using a quantum computer to process and transform quantum data into a form that classical ML algorithms can understand \cite{Chalumuri2021}. The second method transfers data, like density matrix, from the quantum to the classical domain \cite{Fanizza2022}, which is being addressed in the following subsection. Both methods are in the early stages of research but have the potential to offer significant benefits.

\subsection{Density Matrix and Entanglement}\label{Sec.2.1}
Multi-party quantum systems exhibit quantum entanglement between qubits. The density matrix is a Hermitian matrix that allows us to expand its states and probabilities. Generally, it is defined as
\begin{equation}
\rho=\sum\nolimits_{i=1}^{N} \lambda_i|\Psi_i\rangle\langle\Psi_i| ,
\label{eq0}
\end{equation}
where, $\sum_{i}\lambda_i=1$
For all states $|\Psi\rangle$. It represents a multi-party quantum system, especially when it is in a mixed state. So, they allow us to calculate properties such as entanglement and coherence of an actual quantum system and obtain their expectation values and time evolution \cite{paini2021, gu2019}. Hence, the equation for the expectation value in pure states given as $\langle \hat{A}\rangle=\langle\psi|\hat{A}|\psi\rangle$, turns into $\langle \hat{A}\rangle= \mathrm{Tr}(\hat{\rho}\hat{A})$. Similarly, the evolution equation for pure states described as $\frac{d}{dt}|\psi(t)\rangle=\frac{1}{i\hbar}\hat{H}(t)|\psi(t)\rangle$, is transformed into $\frac{d}{dt}\hat{\rho}(t)=\frac{1}{i\hbar}[\hat{H}(t),\hat{\rho}]$, where the relative density matrix of $|\psi\rangle$ is represented by $\hat{\rho}$.

In this study, density matrices are used as data to enable CNNs to extract features from quantum systems. We labeled the data based on the EoF, which for $\rho$ of a system is calculated:
\begin{equation}
    EoF = \inf{\left[\sum_{i}p_iS(\rho^i)\right]},
\label{eq1}
\end{equation}
where the infimum is taken over all possible state decompositions into a probabilistic mixture of pure product states. The von Neumann entropy of the reduced density matrix for subsystem $A$ is calculated by:
\begin{equation}
    S(\rho^A)=-\mathrm{Tr}\left[\rho^A\log_2{(\rho^A)}\right],
\end{equation}
where $\rho^A$ represents the partial trace of density matrix $\rho$ regarding the other subsystem i.e., subsystem $B$.

\subsection{Constructing Extended-Tensor}\label{Sec.2.2}
 QI relies on complex numbers, but ordinary CNNs cannot accommodate them. We convert the density matrix into a three-dimensional tensor, which is less voluminous than regular storage of the complex number  $a +\text{i}\,b$ to the $2\times2$ matrix of:
\begin{equation*}
    \begin{bmatrix}
    a & -b \\
    b & a 
    \end{bmatrix}.
\end{equation*}

\begin{figure}[t]
\centering
\includegraphics[width=1\textwidth]{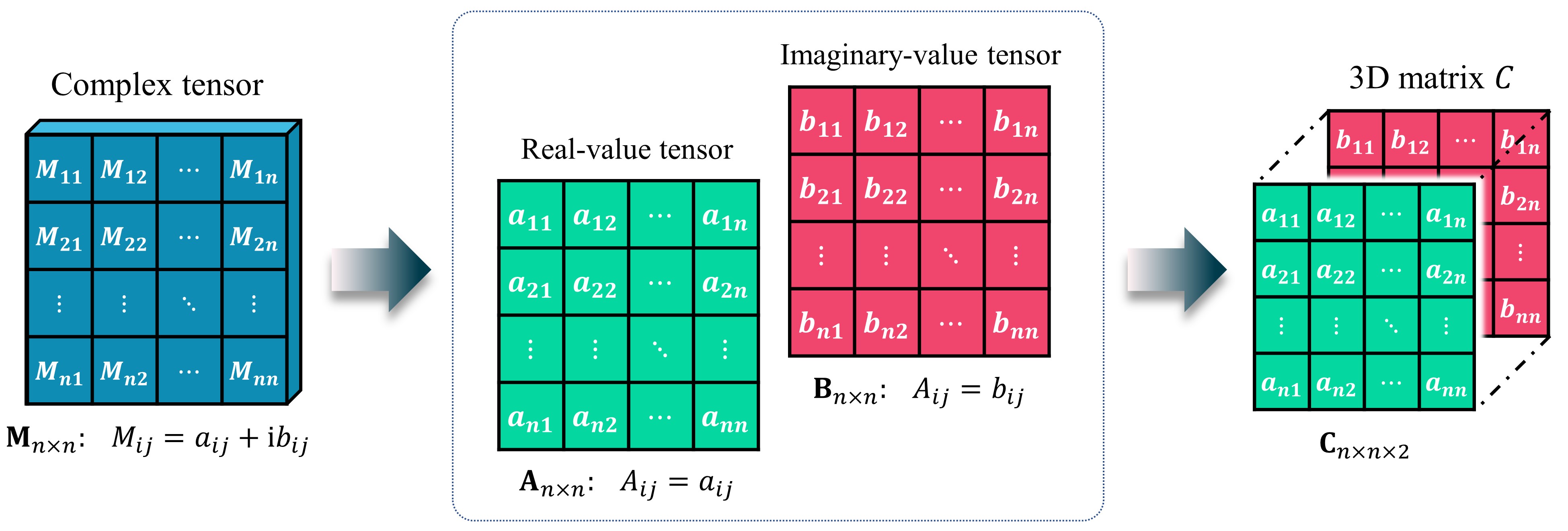}
\caption{The process of converting complex number data to real number data (extended data).}\label{Fig.3}
\end{figure}

Let $\mathbf{M}_{n\times n}$ be the initial matrix made up of complex numbers and $M_{ij}=a_{ij}+\text{i}\,b_{ij}$ be defined as the complex number in the row $i^\text{th}$ and the column $j^\text{th}$ of $\mathbf{M}_{n\times n}$. 
As illustrated in Fig. \ref{Fig.3}, we separate the density matrices into two matrices $\mathbf{A}_{n\times n}$ and $\mathbf{B}_{n\times n}$. Tensor $\mathbf{A}_{n\times n}$ contains real part coefficient,
\begin{equation}
	A_{ij}=\Re(M_{ij})=a_{ij}\ \ \  \text{for} \ \ \ i,j = 1, 2, ..., n
\end{equation}
and tensor $\mathbf{B}_{n\times n}$ contains imaginary part coefficient,
\begin{equation}
	B_{ij}=\Im(M_{ij})=b_{ij}\ \ \  \text{for} \ \ \ i,j = 1, 2, ..., n
\end{equation}
Finally, we obtain a 3D order tensor that forms the input data as follows:
\begin{equation}
	\begin{cases}
		C_{ij0} = A_{ij}&\ \ \    \text{for}\ \ \   i,j = 1, 2, ..., n\\
		C_{ij1} = B_{ij}&\ \ \    \text{for}\ \ \   i,j = 1, 2, ..., n
	\end{cases}
\end{equation}
Tensor $\mathbf{C}_{n\times n\times 2}$ congregates matrix $\mathbf{A}_{n\times n}$ and $\mathbf{B}_{n\times n}$, which have separated the real and imaginary parts into different layers of the 3D on each row $i$ and column $j$.

\subsection{Two-qubit data generation}\label{Sec.2.3}
 To generate the density matrix for a two-qubit system, we generate a random complex matrix $\mathbf{M}_{4 \times 4}$ with elements of $M_{i,j} = a_{ij}+\text{i}\,b_{ij}$ where $a_{ij}$ and $b_{ij}$ are drawn from a Gaussian distribution. A Hermitian matrix $H$ is formed from the random complex matrix $\mathbf{M}_{4\times 4}$ by calculating $H = \mathbf{M} + \mathbf{M}^\dagger$ to ensure that $H$ is Hermitian. Then $\text{Tr}(H^2)>0$ is checked to ensure $\langle\psi|\rho| \psi\rangle\geq0$. Then the density matrix $\rho$ is formed from the Hermitian matrix $H$ by normalizing it, i.e., $\rho = H / \text{Tr}(H)$, to ensure that the $\text{Tr}(\rho)=1$. QuTiP facilitates this procedure \cite{qutip2}.  
We regulate our density matrices to have 75\% non-zero elements so it is more likely to generate a complex, mixed quantum state and feature significant quantum superposition. However, since begetting random two-entangled $\rho$ is approximately three times more likely than begetting separable $\rho$, it is vital to balance our dataset to prevent bias and improve CNN's accuracy.

We generate one million random density matrices as our dataset \cite{alikookani_2022}. This dataset contains 500,000 entangled and 500,000 separable $\rho$ that are explained in  Appendix \ref{Appendix.A} . The matrices are labeled by calculating the EoF of each matrix using the Qiskit \cite{Qiskit}.
\begin{equation}
    \text{Label}(\rho) = 
    \begin{cases} 
    \text{Ent.}, & \text{EoF}(\rho) > 0 \\
    \text{Sep}, & \text{EoF}(\rho) = 0 
    \end{cases}
\end{equation}
The amount of entanglement is also stored thanks to the EoF function.

\subsection{Three-qubit data generation}\label{Sec.2.4}

\begin{figure}[t]
\centering
\includegraphics[width=\textwidth]{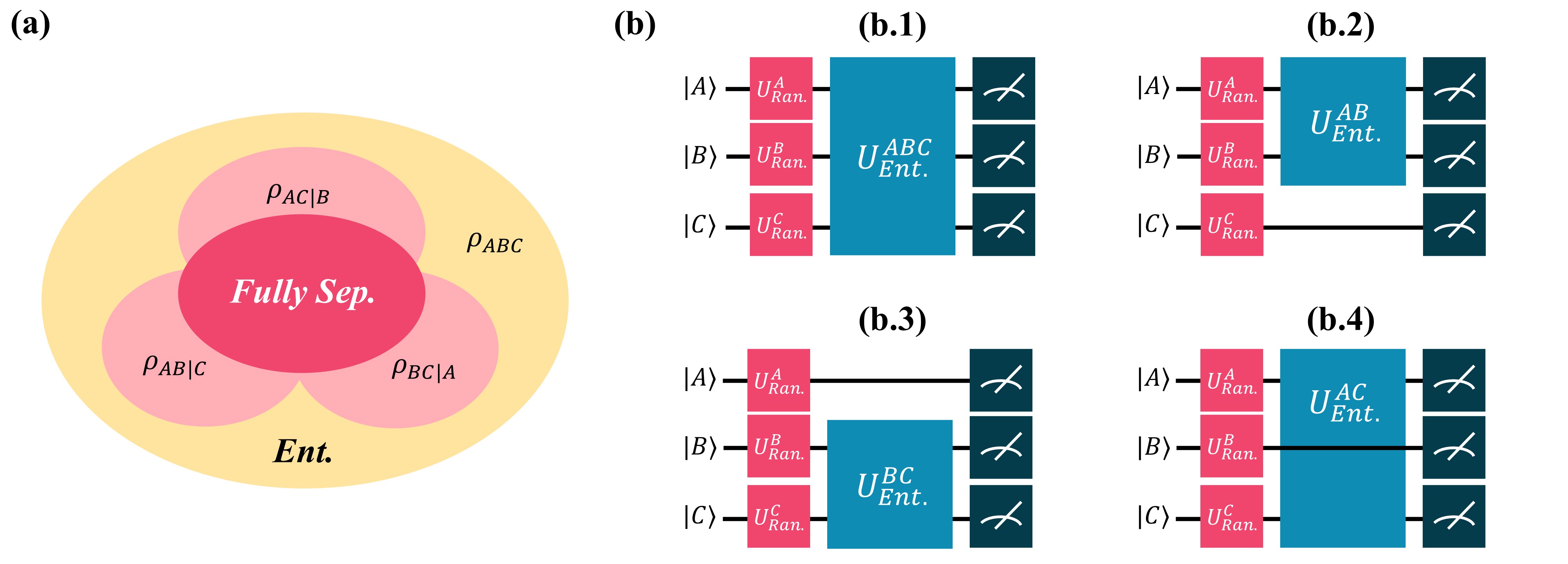}
\caption{An overview of three-qubit: (a) An allegory of dividing state space into categories. (b) Quantum circuits yielding distinct entangled systems: (b.1) Fully entangled state $\rho_{ABC}$, (b.2) Partial entangled states $\rho_{AB|C}$, (b.3) Partial entangled states $\rho_{BC|A}$, (b.4) Partial entangled states $\rho_{AC|B}$.}\label{Fig.4}
\end{figure}

In the three-qubit quantum system Hilbert space onwards, it is not pervasive to only divide states into entangled or separate. As depicted in Fig. \ref{Fig.4}(a), entangled states itself are divided into four types $|\Psi_{ABC}\rangle$, $|\Psi_{AB|C}\rangle$, $|\Psi_{BC|A}\rangle$, and $|\Psi_{AC|B}\rangle$. We detailed in  Appendix \ref{Appendix.B}  how to obtain each type but superficially saying, partial entanglements $|\Psi_{AB|C}\rangle$, and $|\Psi_{BC|A}\rangle$ are provided with a tensor operator ($\otimes$) between an entangled $|\Psi_{i,j}\rangle$ and $|\Psi_k\rangle$ but $|\Psi_{AC|B}\rangle$ is not that easy. Also, $|\Psi_{ABC}\rangle$ are demanding, which puts our focus on known three-qubit states: GHZ state, W state, and Graph state. Fig. \ref{Fig.4}(b) physically demonstrates the entangled states' type, which exerts randomness by single random unitary operations  $U_{Ran.}^{A, B, C}$  and entanglement by $U^{Q}_{Ent.}$ operator, where $Q$ determines monolith entangled qubits. The number of $Q$ over $N$ qubits is as follows:
\begin{equation}
q = \sum_{i=2}^N \binom{N}{i},
\label{eq2}
\end{equation}
when $N=3$ qubit, $q$ is 4. $Q$ includes \{$AB|C$, $BC|A$, $AC|B$, $ABC$\}.

We create 250,000 density matrices for three-qubit states, divided into five balanced categories as a dataset, which takes only 102 seconds to prepare states with purity of 100\% when running on a regular CPU.

\section{Designing a CNN}\label{Sec.3}
CNNs consist of convolutional layers with multiple kernels that extract patterns from low to high-level features, allowing complicated patterns within a matrix to be extracted.

In CNN, the 3D convolution operation is expressed as:
\begin{equation}
    Y_{i,j,k}=\sigma\left(b+\sum_{p=0}^{P-1}\sum_{q=0}^{Q-1}\sum_{r=0}^{R-1}X_{i+p,j+q,k+r}\times W_{p,q,r}\right),
    \label{eq3}
\end{equation}
where $Y_{i,j,k}$ is the output element at position $(i,j,k)$, $X_{i+p,j+q,k+r}$ is the input element at position $(i+p,j+q,k+r)$, $W_{p,q,r}$ is the weight parameter at position $\left(p,q,r\right)$, $b$ is the bias term, and $\sigma$ is the activation function. In equation (\ref{eq3}), $P$, $Q$, and $R$ are the kernel dimensions that work as a filter. The sum over $p$, $q$, and $r$ represents the convolution operation, where the kernel slides over the input tensor and filters the input tensor's corresponding elements. The activation function introduces non-linearity into the layers, allowing the network to learn complex patterns in high-dimensional space.

Entanglement detection requires a high-capacity CNN with many layers. However, longer sequences of layers increase the chance of vanishing gradients of back-propagation. This is caused by activation functions that map input values to small intervals. Leaky ReLU, defined as:
\begin{equation}
    f(x) = 
    \begin{cases}
      ax, & x<0 \\
      x, & \text{else}
    \end{cases},
\end{equation}
is a solution to vanishing gradients as it introduces a slight negative slope $a$ for values below zero, allowing for continued learning.

To understand CNN performance, a loss function is needed.  In a classification task,  the loss function categorical cross entropy (CCE) is calculated for a given number of classes, denoted as $C$.
\begin{equation}
    L_{CCE}=-\sum_{i=1}^{C}{y_i\log{(\hat{y_i})}}
\label{eq4}
\end{equation}
In CCE's Equation (\ref{eq4}), $y_i$ represents the actual label for class $i$, and $\hat{y_i}$ denotes the predicted SoftMax probability for class $i$. In 2-qubit system, considering only entangled or not, the proposed equation (\ref{eq1}) simplifies to a binary classification problem by setting $C=2$. The modified equation becomes:
\begin{equation}
    -\sum_{i=1}^{2}y_i\log{(\hat{y_i})}=-y_1\log{(\hat{y_1})}-y_2\log{(\hat{y_2})}
    \label{eq5}
\end{equation}
Equation (\ref{eq5}) is further simplified to represent the formula for binary cross entropy as follows:
\begin{equation}
    L_{BCE}=-y\log{(\hat{y})}-(1-y)\log{(1-\hat{y})}
\end{equation}
Furthermore, the model functions for regression task to estimate the EoF value of each density matrix can be define by shifting the loss function to mean square error (MSE), which is obtained as follows:
\begin{equation}
L_{MSE} = \frac{1}{N_s} \sum_{i=1}^{N_s} (y_i - \hat{y}_i)^2
\end{equation}
where $N_s$ is the total number of observations, $y_i$ is the EoF of the $i^{th}$ instance, and $\hat{y}_i$ is the corresponding model prediction.

The back-propagation algorithm optimizes weight and bias in equation (\ref{eq3}) over loss functions $L$ as follows:
\begin{equation}
\theta^{k-1} = \theta^k - \alpha \frac{\partial L}{\partial \theta^k}
\end{equation}
Here, $\theta^k$ represents the parameters at layer $k$.

A CNN is designed and developed by incorporating and assembling these components \cite{chollet2021learning}.

\subsection{Proposed CNN}\label{Sec.3.1}
\begin{figure}[t]
\centering
\includegraphics[width=\textwidth]{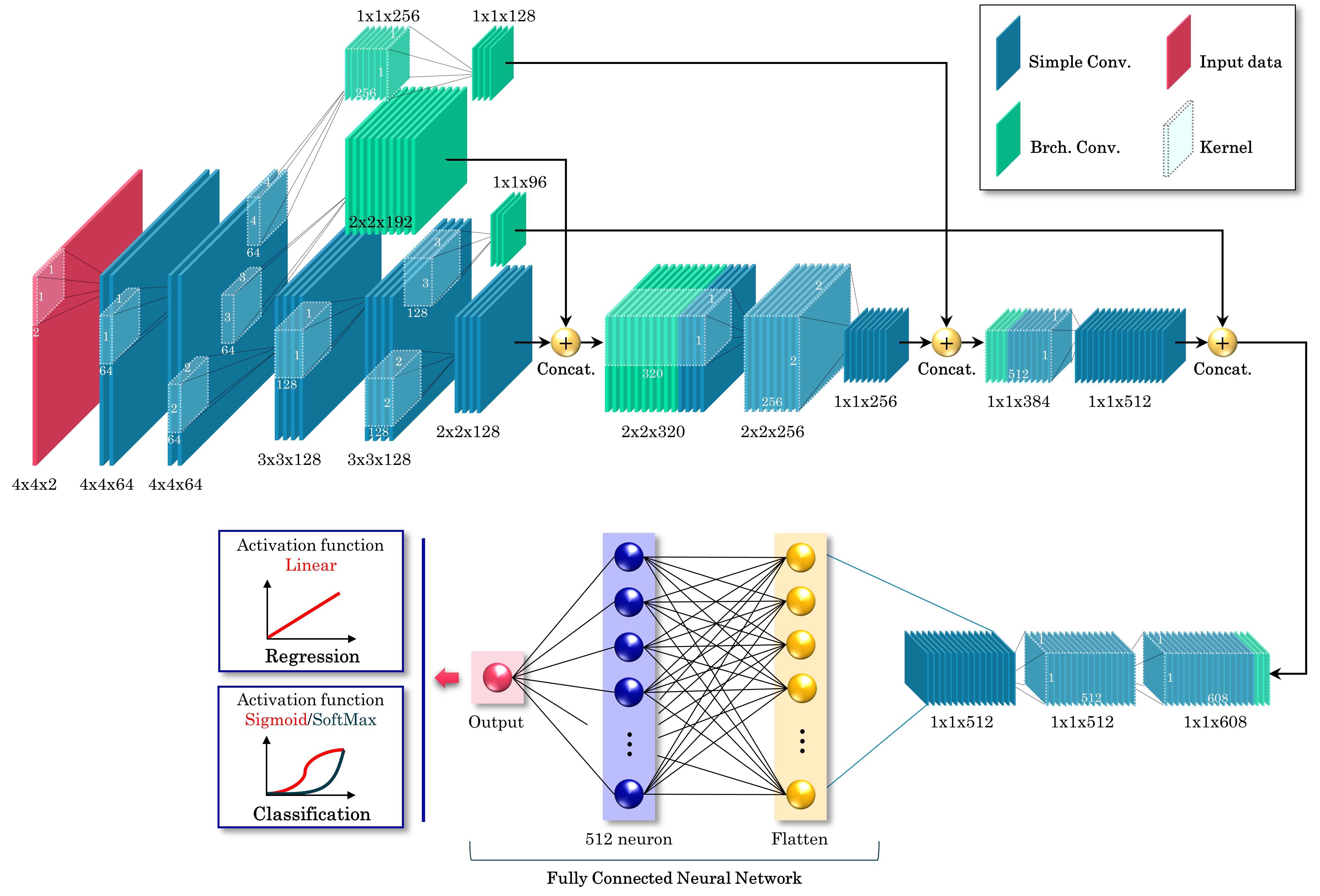}
\caption{XpookyNet's whole scheme. Blue layers show the principal artery and green layers show the branches. The last layer consists of two fully connected dense layers.}\label{Fig.5}
\end{figure}

CNN requires layers to be appropriately arranged and select hyper-parameters wisely, such as the number of kernels and filters in each layer. We set different kernel sizes, as shown in Fig. \ref{Fig.5}. Kernel size selection depends on the area it covers in the input tensor. Smaller kernels, e.g., $2\times2$ detect detailed patterns and larger kernels, e.g., $4\times4$ detect broader patterns \cite{jia2022}.

The phenomenon of entanglement, often referred to as "\textit{spooky action at a distance}", is central to this study, which introduces a CNN framework named XpookyNet. This framework is designed to identify entangled states and quantify their degree by employing various optimization and learning techniques. Figure \ref{Fig.5} depicts the entire XpookyNet architecture, which incorporates several enhancement methods. The core structure of XpookyNet comprises 10 convolutional layers (Simple Conv.) and two fully connected neural networks (NN). The enhancement methods of XpookyNet include (1) batch normalization combined with separable convolutional layers (B.N. Sep.), (2) three-branch convolutional layers (Brch.) and (3) the learning rate reduction method on plateaus (Plat.).

In the context of batch normalization with separable convolutional layers, batch normalization improves the stability, performance, accuracy, and speed of the CNN \cite{santurkar2018}. In addition, separable convolutional layers make CNN training more efficient, faster, and precise by lowering computational complexity and mitigating overfitting. However, they might be less effective with low-dimensional tensors and typically necessitate more layers to achieve the desired accuracy \cite{liu2020}.

\subsection{Training CNNs}\label{Sec.3.3}
XpookyNet uses the stochastic gradient descent (SGD) optimizer with momentum and saves the most accurate model with minimum loss. The learning rate is adjusted when the loss reaches a plateau \cite{Yoshida2020Data-dependence}. Parallel paths help prevent vanishing gradients as well by giving a shortcut to backward flow \cite{Lan2019Image}. Also, balancing and shuffling the data improves results.

When training models, it is crucial to select the learning rate carefully. The learning rate determines how much the model changes in response to the error estimate. It is recommended to begin with a high learning rate as it enables the model to learn faster. The learning rate is reduced manually when the loss function stops improving, allowing the model to make more miniature adjustments. This process is repeated throughout the training to guide the model towards better performance \cite{Givi2015Modelling, Yu2020LLR}.

\section{Results Evaluation}\label{Sec.4}
We assess XpookyNet with different structures for detecting entanglement in quantum systems. Our analysis includes absolute error and accuracy plots, overview tables, and classification and regression performance using relevant metrics.

Accuracy (ACC) measures the ratio of correctly predicted instances to the total number of instances in the dataset.
\begin{equation}
    ACC=\frac{TP+TN}{TP+TN+FP+FN}
\end{equation}
where True Positives ($TP$) is the number of correctly predicted positive instances, True Negatives ($TN$) is the number of correctly predicted negative instances, False Positives ($FP$) is the number of incorrectly predicted positive but are negative, False Negatives ($FN$) is the number of incorrectly predicted negative but are positive.

Mean absolute error (MAE) measures the average magnitude of errors between predicted values and actual values to asses the regression performance.
\begin{equation}
    MAE = \frac{1}{N_s} \sum_{i=1}^{N_s} \left| y_i - \hat{y}_i \right|
\end{equation}
where $N_s$ is the total number of instances. $y_i$ represents the actual value of the $i^{th}$ instance. $\hat{y}_i$ represents the predicted value for the $i^{th}$ instance, and $y_i$ is the actual value to asses the regression performance.

False Negative Rate (FNR) measures the proportion of actual positive instances that were incorrectly predicted as negative.
\begin{equation}
    FNR = \frac{FN}{TP+FN}
\end{equation}

Matthews Correlation Coefficient (MCC) is a balanced measure that takes into account true and false positives and negatives.
\begin{equation}
    MCC = \frac{TP\times TN-FP\times FN}{\sqrt{(TP+FP)(TP+FN)(TN+FP)(TN+FN)}}
\end{equation}

\subsection{Two-qubit Entanglement Detection}
Results of 10,000 random two-qubit test states are presented in Table \ref{Table.1} for entanglement detection in two qubits, which is crucial for building and characterizing larger quantum systems and their states \cite{Virz2019}.

\begin{table*}[t]
    \footnotesize
    \centering
    \caption{
    Overview of XpookyNet's performance with different structures used in this study for two-qubit entanglement detection.
    }
    \begin{tabular}{p{0.21\textwidth} p{0.09\textwidth}p{0.09\textwidth}p{0.09\textwidth} p{0.11\textwidth} p{0.11\textwidth} p{0.11\textwidth}}
\hline\\[-1ex] 
&&&&&\multicolumn{2}{c}{Model detail} \\[+1ex] 
\\[-1ex] 
Structure&\centering ACC&\centering FNR&\centering MCC&\centering Epoch time&Number of Conv. layers&Number of parameters\\[+1ex] 
\hline
\hline\\[-1ex] 
NN                 &\centering 0.8325&\centering 0.1826&\centering 0.6603&\centering 19s44ms&\centering0&3,233\\
Simple Conv.       &\centering 0.9500&\centering 0.0622&\centering 0.9189&\centering 51s55ms&\centering10&1,334,913\\
Brch.              &\centering 0.9624&\centering 0.0377&\centering 0.9276&\centering 63s22ms&\centering14&2,015,521\\
B.N. Sep.           &\centering 0.9632&\centering 0.0504&\centering 0.9205&\centering 57s20ms&\centering10&1,080,263\\
Plat.              &\centering 0.9789&\centering 0.0244&\centering0.9576&\centering 48s33ms&\centering10&1,334,913\\
\textbf{Brch. Plat.}         &\centering \textbf{0.9852}&\centering \textbf{0.0156}&\centering \textbf{0.9712}&\centering \textbf{64s58ms}&\centering\textbf{14}&\textbf{2,015,521}\\
Plat. B.N. Sep.      &\centering 0.9749&\centering 0.0440&\centering 0.9452&\centering 60s47ms&\centering10&1,080,263\\
Brch. B.N. Sep.     &\centering 0.9664&\centering 0.0518&\centering 0.9315&\centering 75s62ms&\centering14&1,324,615\\
Brch. Plat. B.N. Sep.&\centering 0.9761&\centering0.0440&\centering0.9452&\centering 75s95ms&\centering 14&1,335,430\\
\hline
    \end{tabular}
    \label{Table.1}
\end{table*}

Strategic model selection and combination are crucial for optimal performance. Combining models can lead to continued improvement, emphasizing the need for a nuanced approach to model architecture and configuration.

The NN is a baseline with the lowest ACC, MCC, and a notable FNR. The structure Simple Conv. improves ACC and MCC but retains a discernible FNR. Three branch layers (Brch.) enhances accuracy but results in a higher FNR than Simple Conv., albeit with an improved MCC. B.N. Sep. achieve a commendable balance with high ACC, a lower FNR, and an impressive MCC. The Reducing the learning rate on plateaus (Plat.) outperforms all others with the highest ACC, MCC, and minimal FNR. Combining branched CNN with the Plat. method (Brch. Plat.) is the most optimal structure.

\begin{figure}[t]
\centering
\includegraphics[width=1\textwidth]{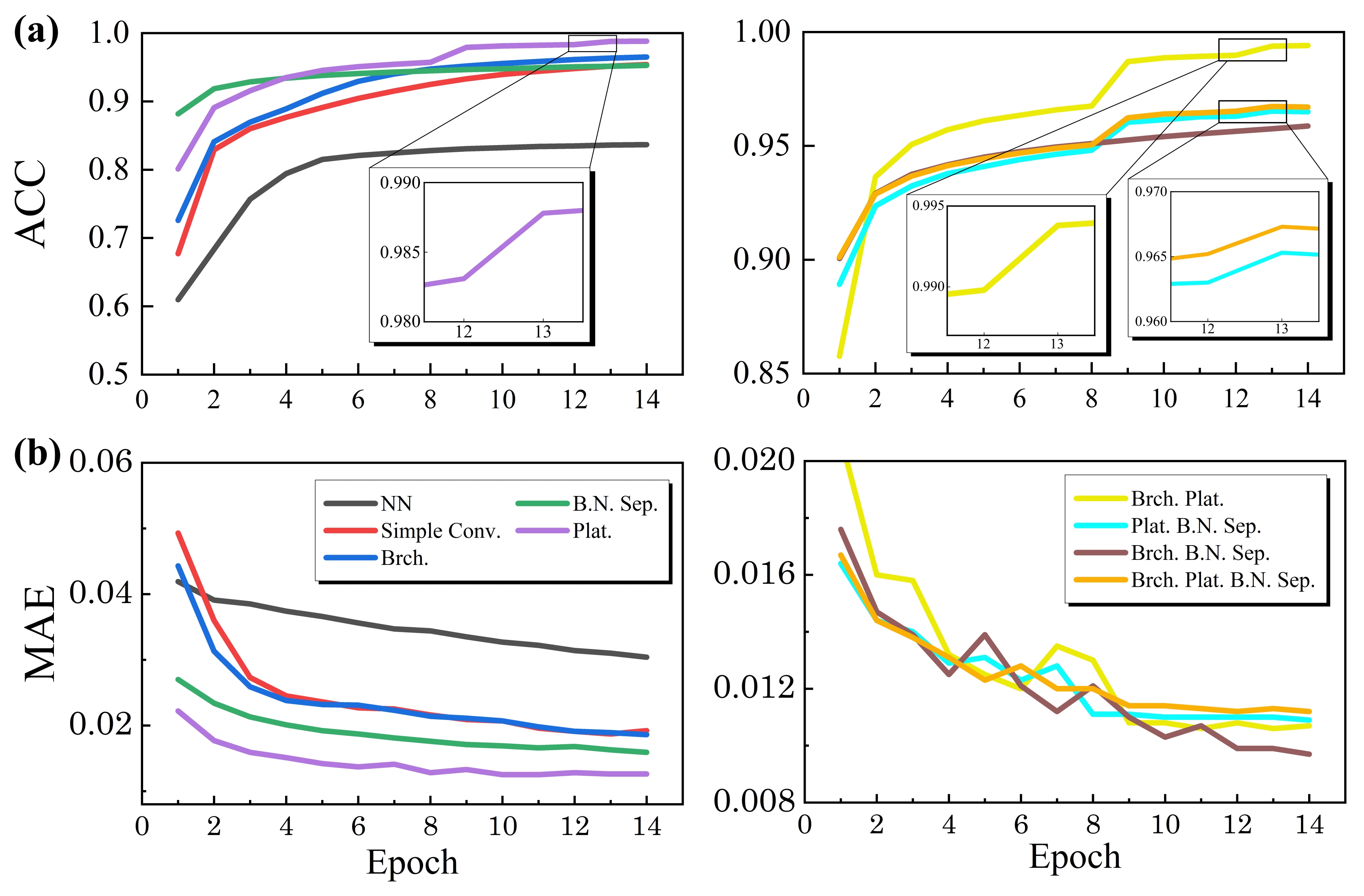}
\caption{The proposed XpookyNet's performance in the learning process. The progression of each XpookyNet structure's performance metrics on (a) the classification, and (b) the regression tasks.}\label{Fig.6}
\end{figure}

In only 14 epochs, XpookyNet achieves 98.53\% accuracy. The network can even determine the entanglement degree EoF with a MAE of 0.0095.
Fig. \ref{Fig.6}(a) depict the performance evolution of methods over epochs. The NN starts with high accuracy but flats around ACC $\approx0.8$. In comparison, convolutional models show progressive improvement, especially the ones that exploit Plat. technique. Altogether, the combined structure Brch. Plat. stands out for its significant improvement over time.
Besides, the effect of Plat. method on epochs $8$ and $12$ is showcased.
Fig. \ref{Fig.6}(b) also compares the decrease of MAE under each method and their combination, pointing out how combining the methods leads to a lower error in estimating the EoF of the states.

\subsection{Incomplete Density Matrix}\label{Sec.5.2}
Quantum tomography is a procedure used to determine the complete quantum state of a system. The primary outcome of tomography is the density matrix, which captures a quantum system's statistical properties and provides a full description of the quantum state. However, tomography is notoriously costly in terms of time and resources due to the exponential growth in complexity with the size of the system \cite{Lanyon2016Efficient}.

In this subsection, we explore handling incomplete density matrices to evaluate model performance on fewer measurements. We consider the primary dataset described in subsection \ref{Sec.2.3} as a training dataset, generated from complete density matrices with full tomography. By relying only on machine learning capacity, this study investigates the possible elimination of the need for costly full tomography, evaluating the performance of XpookyNet in classifying incomplete data.

In quantum tomography, the density matrix is reconstructed by preparing multiple identical copies of the quantum state, in which measurements are repeated on multiple bases that span the whole state space. The measurement operators, which are the basis for an $N$-qubit system, are constructed as tensor products of Pauli matrices ($\sigma_1, \sigma_2, \sigma_3$), along with the identity matrix ($\sigma_0$), resulting in a set of $2^N\times 2^N=4^N$ different operators. The linear combination of these operators assembles the density matrix as follows:
\begin{equation}
    \rho = \frac{1}{2^N} \sum_{\Bar{v}} \mathrm{Tr}(\sigma_{v_1}\otimes \sigma_{v_2}\otimes \cdots \otimes\sigma_{v_N} \rho)\sigma_{v_1}\otimes \sigma_{v_2}\otimes \cdots \sigma_{v_N}
    \label{eq20}
\end{equation}
The summation encompasses vectors $\Bar{v}=(v_1,\cdots,v_N)$, where each entry $v_i$ is selected from the set $i=0, 1, 2, 3$. By conducting measurements of observables that are products of Pauli matrices, we can estimate each component within this summation, thereby enabling us to approximate $\rho$ \cite{nielsen2001quantum}.

In evaluation process, we generate incomplete data from complete data so that XpookyNet's performance can be assessed using fewer measurements. While a complete density matrix $\rho$ includes all measurement bases $\Bar{v}$, an incomplete density matrix comprises only a subset of the total bases. Assuming $\Bar{v'}$ are the arbitrary ignored measurement bases in the quantum tomography process, the incomplete density matrix $\rho'$ can be generated from the complete density matrix as follows:

\begin{equation}
    \rho^{'} = \rho-\frac{1}{\mathcal{N}} \sum_{\Bar{v'}} \mathrm{Tr}(\sigma_{v'_1}\otimes \sigma_{v'_2}\otimes \cdots \otimes\sigma_{v'_N} \rho)\sigma_{v'_1}\otimes \sigma_{v'_2}\otimes \cdots \sigma_{v'_N}
    \label{eq21}
\end{equation}

In a 2-qubit system, we evaluate 10,000 instances of incomplete test data generated from complete test data according to equation (\ref{eq21}). Incomplete data are characterized by a random subset of measurement bases, with varying numbers of ignored measurement bases. 

\begin{figure}[t]
\includegraphics[width=\textwidth]{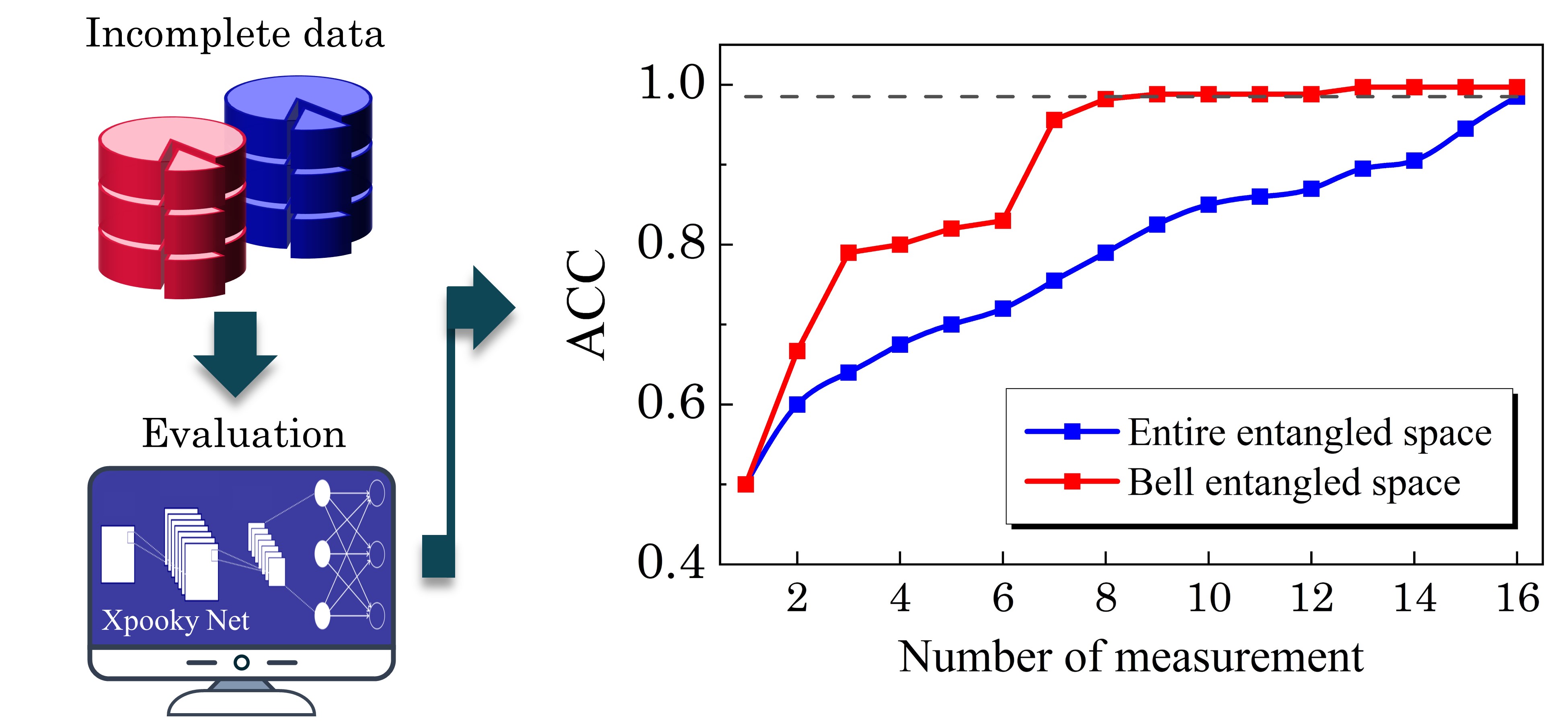}
\caption{Accuracy plot of  XpookyNet's performance for the classification task on incomplete data in two different data spaces: Entire entangled space and Bell entangled space (The dashed line represents the best model accuracy in Table \ref{Table.1} on complete data: ACC = 0.9852).}
\label{Fig.7}
\end{figure}

As depicted in Fig. \ref{Fig.7}, XpookyNet's performance is a trade-off for the number of measurements with accuracy, differing between entire and Bell entangled spaces. In the entire entangled space, the model maintains around 80\% accuracy with half the density matrix elements. Detecting Bell entangled states with fewer measurements is crucial for efficient verification, enhancing the feasibility and scalability of quantum information processing. In Bell entangled space, accuracy remains above 90\% with seven measurements and above 80\% with only three measurements.

\subsection{Three-qubit Entanglement Detection}\label{Sec.5.3}
Improving XpookyNet becomes essential, especially for analyzing quantum states at different purity levels to address noise, errors, and imperfections in the NISQ era.
Hence, XpookyNet utilizes B.N. Sep. method in a three-qubit setting for enhanced performance on larger tensors compared to traditional convolutional layers. 

\begin{figure}[t]
\includegraphics[width=\textwidth]{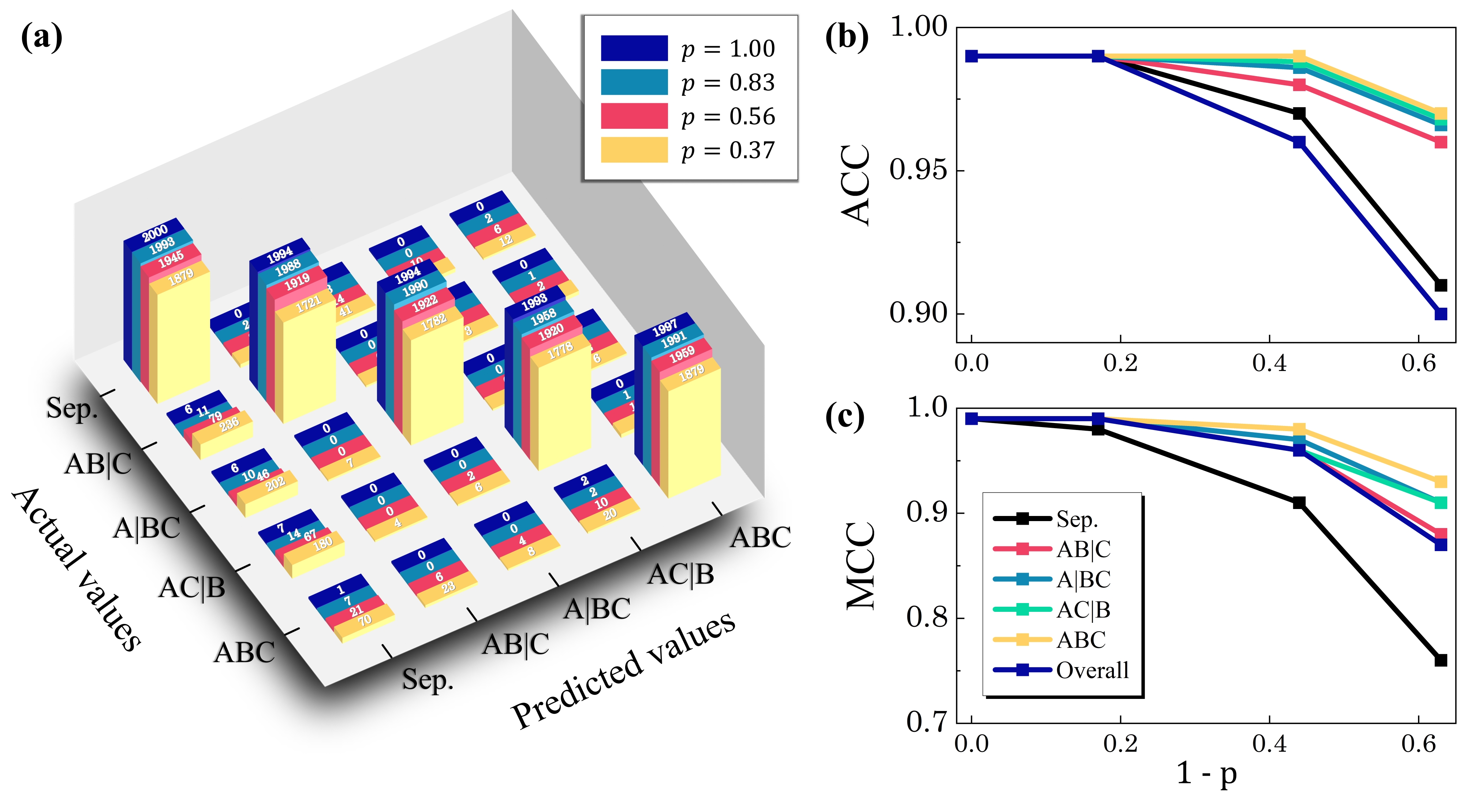}
\caption{Based on purity values of $p=1, 0.83, 0.56, 0.37$, (a) confusion matrices, (b) accuracy plot, and (c) matthews correlation coefficient plot illustrate the classification performance of XpookyNet in three-qubit quantum systems.}\label{Fig.8}
\end{figure}

XpookyNet ("Brch. Plat. B.N. Sep. structure") classifies three-qubit systems with high accuracy on density matrices with pure state $p=1$, but its performance declines as purity decreases, depicted in Fig. \ref{Fig.8}. Real-world quantum systems yield less coherence due to imperfections and noise; decoherent systems induce decreased density matrix purity. Since decoherence disrupts quantum correlations within the quantum system, quantum systems lose entanglement as purity reduces \cite{essakhi2022intrinsic}. Hence, the purity plunge primarily turns down XpookyNet's functionality in partial entangled states encompassing the characteristics of both separable and entangled states, with an average of 11.7\%.
As purity decreases, fully entangled states can appear partial entangled or separable, leading to misclassification in XpookyNet. Mixing density matrices can also blur class boundaries. However, XpookyNet performs strongly, with only a 5.98\% performance decrease on average, despite purity drop effects in these two classes.
This behavior of XpookyNet aligns with the physical intuition about quantum entanglement and the effects of system imperfections, demonstrating the network's ability to capture the complex dynamics of quantum systems despite noise and imperfections \cite{Facchi2019Phase}.

\begin{table}[t]
    \footnotesize
    \centering
    \caption{Overview of metrics used to assess three-qubit entanglement detection.}
\begin{tabular}{|c|cccc|cccc|cccc|}
\hline
& \multicolumn{4}{|c}{} & \multicolumn{4}{|c}{} & \multicolumn{4}{|c|}{}   \\ 
& \multicolumn{4}{|c}{\textbf{ACC}} & \multicolumn{4}{|c}{\textbf{FNR}} & \multicolumn{4}{|c|}{\textbf{MCC}} \\[+1.5ex] \hline
\multirow{2}{*}{\textbf{Class/p}}&\multirow{2}{*}{1}&\multirow{2}{*}{0.83}&\multirow{2}{*}{0.56}&\multirow{2}{*}{0.37}&\multirow{2}{*}{1}&\multirow{2}{*}{0.83}&\multirow{2}{*}{0.56}&\multirow{2}{*}{0.37}&\multirow{2}{*}{1}&\multirow{2}{*}{0.83}&\multirow{2}{*}{0.56}&\multirow{2}{*}{0.37} \\
&&&&&&&&&&&&   \\ \hline
&&&&&&&&&&&&   \\ 
\centering Sep.&0.99&0.99&0.97&0.91&0.00&0.00&0.02&0.06&0.99&0.98&0.91&0.76\\[+1.5ex]
\centering $AB|C$&0.99&0.99&0.98&0.96&0.00&0.00&0.04&0.13&0.99&0.99&0.96&0.88\\[+1.5ex]
\centering $A|BC$&0.99&0.99&0.99&0.97&0.00&0.00&0.02&0.10&0.99&0.99&0.97&0.91\\[+1.5ex]
\centering $AC|B$&0.99&0.99&0.99&0.97&0.00&0.00&0.04&0.11&0.99&0.99&0.96&0.91\\[+1.5ex]
\centering $ABC$&0.99&0.99&0.99&0.97&0.00&0.00&0.02&0.06&0.99&0.99&0.98&0.93\\[+1.5ex] \hline
&&&&&&&&&&&&\\
\centering\textbf{overall}&0.99&0.99&0.96&0.90&0.00&0.00&0.03&0.09&0.99&0.99&0.96&0.87\\[+1.5ex] \hline
\end{tabular}
    \label{Table.2}
\end{table}

XpoolyNet's performance declines across all classes as purity decreases from 1 to 0.37. Table \ref{Table.2} expresses ACC and MCC decrease while FNR increases, indicating misclassifications.
For separable states, purity decline results in a 7.9\% ACC decrease and a 23.4\% MCC drop. The FNR increases by 0.0605, a notable rise in errors, especially FP. TN and TP remain relatively accurate, pointing to entanglement features obscuring results.
As illustrated in Fig. \ref{Fig.8},  partial entangled states ($|\psi_{AB|C}\rangle$, $|\psi_{A|BC}\rangle$, $|\psi_{AC|B}\rangle$) show less drastic ACC and MCC decrease than separable states. However, the FNR increases by an average of 62.5\%, indicating significant errors at decimated purity.
Fully Entangled ($|\psi_{ABC}\rangle$) states experienced a negligible impact on purity decrease, with only a 1.83\% ACC drop and a 6.36\% MCC decrease. Oddly, three-qubit full entanglement detection functions better than two-qubit problem due to the limited known tripartite entangled set of GHZ, W, and graph states.

\section{Discussion and Conclusion}\label{Sec.5}
In the quest to advance quantum computing, entanglement detection stands as a challenge. Numerous researchers have embarked on this endeavour, employing network-based ML techniques to achieve higher accuracy and productivity. Several contributions are compared here, highlighting their unique methodologies and findings and showing how ours advances them.

M. Yosefpor et al. leverage autoencoder NNs to optimize entanglement detection by identifying a minimal set of measurements for detecting entangled states, reducing experimental costs while achieving the high accuracy of 98\% with full tomography \cite{Yosefpor_2020}. The research by Y. Chen et al.'s method for entanglement detection uses unsupervised learning \cite{ chen2021}. Their approach leverages the convexity of non-entangled samples to design a complex-valued network, combining a pseudo-siamese network with a generative adversarial network. This methodology, trained with separable states, achieves a detection accuracy of over 97.5\%. N. Asif et al. \cite{asif2023} developed an NN model aimed at classifying quantum states by optimizing a Bell-type inequality for the relative entropy of coherence. Their method, which labels quantum states using the PPT criterion, achieves an accuracy of up to 94.62\% and demonstrates scalability to more complex systems. YD. Qu et al. \cite{Qu2023EntanglementDW} focus on using complex-valued CNN to classify quantum states, achieving an average accuracy of 98.7\%, with peak performance at 99.99\% for only four-qubit pure states. The researchers emphasize the importance of incorporating phase and amplitude information in datasets to classify accurately. J. Pawlowski et al. \cite{pawlowski2022quantification} uniquely address the identification of positive-under-partial-transposition entangled states (PPTES) through entanglement-preserving symmetry operations within a triple Siamese network. This semi-supervised approach excels in analyzing specific state classes, including Acin, UPB, Horodecki, mixed, and pure states, demonstrating a high accuracy, particularly in detecting pure states. 

There are, however, difficulties with generalizability due to their reliance on limited states for training, as well as the computational complexity of their architectures, particularly when dealing with real-world quantum systems. Due to the complexity of their models or their sensitivity to boundary data, scaling up to larger quantum systems requires substantial resources and expertise.

The proposed XpookyNet framework offers a streamlined and efficient approach by representing density matrices as 3D tensors for entanglement detection using CNNs. Compared with its counterparts, XpookyNet has a much higher level of accuracy in terms of pervasive evaluation metrics, as it inherits parallel layers and utilizes tailored kernel sizes; it learns data quickly with fewer training epochs due to its devised confrontation with plateauing. 

The versatility of XpookyNet allows it to be used for regression tasks to estimate the EoF of a density matrix. Additionally, our analysis also extends to the impact of decoherence on fully and partially entangled states, considering the purity of density matrices. The model's mastery subsidence with purity drop, in tandem with the growth in classes' errors, shows alignment with physical intuition about quantum entanglement, confirming XpookyNet's connoisseur of quantum idiosyncrasies.
By establishing a procedure that generates all of Hilbert space's partial entangled states $|\psi_{AB|C}\rangle$, $|\psi_{AC|B}\rangle$, and $|\psi_{BC|A}\rangle$ and employing a model that accurately captures the quantum system's characteristics through its density matrix, detecting tripartite entanglement $|\psi_{ABC}\rangle$ becomes straightforward as a multi-classification task. This is because the density matrix of a fully entangled $|\psi_{ABC}\rangle$ system encapsulates the integrals derived from the partial entangled states $|\psi_{AB|C}\rangle$, $|\psi_{AC|B}\rangle$, and $|\psi_{BC|A}\rangle$, paving the way for more qubit entanglement configurations by analogy \cite{xie2021triangle,baccari2017efficient}.
Importantly, our research devises a method to assess the model's performance with fewer measurements, striking a balance between accuracy and the cost of tomography.

\bmhead{Acknowledgements}
No acknowledgments to declare.

\bmhead{Contribution Statement}
\textbf{AK}: Conceptualization, Methodology, Investigation, Programming, Literature survey, Writing. \textbf{YM}: Data Adjustment, Visualization, Literature Survey, Writing. \textbf{PK}: Literature survey, Revising Manuscript. \textbf{HA}: Revising Manuscript, Project administration, supervision. \textbf{KF}: Revising Manuscript, Supervision. \textbf{MB}: Revising Manuscript, Supervision.

\bmhead{Funding}

Not applicable.

\bmhead{Conflict of interest}

Not applicable.

\bmhead{Codes availability}

The codes that support the plots within the paper are publicly available on: \url{https://github.com/AKookani/XpookyNet}

\begin{appendices}

\section{Two-qubit density matrix}\label{Appendix.A}
Two-qubit separable mixed states are expressed:
\begin{equation}
    \rho_{sep}=\sum_{i=1}^{m}{\lambda_i\rho_i^A\otimes\rho_i^B},
\end{equation}
where $\sum_{i}\lambda_i=1$ and $0\le\lambda_i\le1$, summing over $i=1$ to $m$. $\rho_i^A$ and $\rho_i^B$ are arbitrary density matrices of A and B qubits respectively. Two-qubit entangled mixed states are expressed as:
\begin{equation}
    \rho_{ent}=\sum_{i=1}^{m}{\lambda_i\rho_{i}^{AB}},
\end{equation}
 $\rho_{i}^{AB}$ is the arbitrary density matrix of the two-qubit entangled state.

\section{Generation of Three-qubit density matrix}\label{Appendix.B}
In the three qubits case, two-qubit entangled states are required to build partial entangled states. 

\subsection{Three-qubit Separable density matrix}\label{Appendix.B.1}
Separable mixed states are expressed:
\begin{equation}
    \rho_{sep}=\sum_{i=1}^{m}{\lambda_i\rho_i^A\otimes\rho_i^B\otimes\rho_i^C},
\end{equation}
where $\sum_{i}\lambda_i=1$ and $0\le\lambda_i\le1$, with $m$ iterating from 1 to an arbitrary number. The greater the value of $m$, the less pure the state.

\subsection{Partial entangled density matrix}\label{Appendix.B.2}
Partial entangled mixed states of a three-qubit state are produced in three cases.\\
Case 1: Partial entangled pairs $|\psi_{A}\rangle$ and $|\psi_{B}\rangle$ are generated by:
\begin{equation}
    \rho_{AB|C}=\sum_{i=1}^{m}{\lambda_i\rho_i^{AB}\otimes\rho_i^C},
\end{equation}
Where $\sum_{i}\lambda_i=1$ and $0\le\lambda_i\le1$, with $m$ iterating from 1 to an arbitrary number. $\rho^{AB}$ is a density matrix of a two-qubit entangled state.\\
\\
Case 2: Partial entangled pairs $|\psi_{B}\rangle$ and $|\psi_{C}\rangle$ are generated by:
\begin{equation}
    \rho_{A|BC}=\sum_{i=1}^{m}{\lambda_i\rho_i^A\otimes\rho_i^{BC}},
\end{equation}
Where $\sum_{i}\lambda_i=1$ and $0\le\lambda_i\le1$, $\rho_i^{BC}$ is an arbitrary density matrix of a two-qubit entangled state.\\
\\
Case 3: Partial entangled pairs $|\psi_{A}\rangle$ and $|\psi_{C}\rangle$ are generated by state vectors of the entangled state $|\psi_{AC}\rangle$ and separated state $|\psi_B\rangle$ considered as:
\begin{equation}
    |\psi_{AC}\rangle=[a_0\ \ a_1\ \ a_2\ \ a_3]^T
    \label{eqb4}
\end{equation}
and
\begin{equation}
    |\psi_{B}\rangle=[b_0\ \ b_1]^T.
    \label{eqb5}
\end{equation}
Therefore, the partial entangled states are considered as: 
\begin{equation}
|\psi_{AC|B}\rangle=[a_0b_0\ \ a_1b_0\ \ a_0b_1\ \ a_1b_1\ \ a_2b_0\ \ a_3b_0\ \ a_2b_1\ \ a_3b_1]^T,
\end{equation}
where $a$ and $b$ are corespondent amplitudes of states (\ref{eqb4}) and (\ref{eqb5}). Ultimately, the partial entangled states are generated by:
\begin{equation}
    \rho_{AC|B}=\sum_{i=1}^{m}{\lambda_i(|\psi_{B|AC}\rangle_i\langle\psi_{B|AC}|_i)},
\end{equation}
where $\sum_{i}\lambda_i=1$ and $0\le\lambda_i\le1$ and $|\psi_{B|AC}\rangle$ is the generated partial entangled state.

\subsection{Three-qubit GHZ state}\label{Appendix.B.3}
The state vector of a random three-qubit state that GHZ is applied to is described as:
\begin{equation}
    |\Psi_{GHZ}\rangle=\frac{1}{\sqrt{N_{GHZ}}}\left(\cos{(\epsilon)}|000\rangle+\sin{(\epsilon)}e^{i\phi}|\Phi_{ABC}\rangle\right)
\end{equation}
with initial states:
\begin{align}
&|\Phi_{ABC}\rangle = |\varphi_A\rangle|\varphi_B\rangle|\varphi_C\rangle\\
&|\varphi_A\rangle=\cos{(\theta_A)}|0\rangle+e^{i\phi_A}\sin{(\theta_A)}|1\rangle,\\
&|\varphi_B\rangle=\cos{(\theta_B)}|0\rangle+e^{i\phi_B}\sin{(\theta_B)}|1\rangle,\\
&|\varphi_C\rangle=\cos{(\theta_C)}|0\rangle+e^{i\phi_C}\sin{(\theta_C)}|1\rangle,
\end{align} 
where $$N_{GHZ}=1/(1+\cos{(\delta)}\sin{(\delta)}\cos{(\alpha)}\cos{(\beta)}\cos{(\phi)}).$$
The angles belong to the intervals $\delta\in(0,\ \pi/4]$, $(\alpha,\ \beta,\ \gamma)\in(0,\ \pi/2]$, and $\phi\in[0,\ 2\pi)$. \\

\subsection{Three-qubit W-state}\label{Appendix.B.4}
A random state vector that undergoes the W-state operation can be written as:
\begin{align}
\left|\Psi_{\text{W}}\right\rangle \nonumber= &\frac{1}{\sqrt{N_{\text{W}}}} (a\ |001\rangle + b\ |010\rangle+ c |100\rangle - d\ |\phi\rangle),\nonumber \\
&\quad \nonumber \\
\end{align}
where $$N_W=1/\sqrt{|a|^2+|b|^2+|c|^2+|d|^2},$$ is the normalization factor and $|\phi\rangle$ is a superposition of remaining states with W-state.

\subsection{Three-qubit Graph state}\label{Appendix.B.5}
The graph state can be considered as:
\begin{align}
\left|\Psi_{\text{Graph}}\right\rangle \nonumber= &\frac{1}{\sqrt{N_{\text{Graph}}}} (\alpha_0\ |000\rangle + \alpha_1\ |001\rangle + \alpha_2\ |010\rangle - \alpha_3\ |011\rangle \nonumber \\
&\quad  + \alpha_4\ |100\rangle + \alpha_5\ |101\rangle  - \alpha_6\ |110\rangle + \alpha_7\ |111\rangle)\nonumber \\
\label{eq22}
\end{align}
where normalization factor is $$N_{Graph}=1/\sqrt{|\alpha_0|^2+\ldots+|\alpha_7|^2}.$$

\end{appendices}

\bibliography{sn-bibliography}


\begin{thebibliography}{53}
\ifx \bisbn   \undefined \def \bisbn  #1{ISBN #1}\fi
\ifx \binits  \undefined \def \binits#1{#1}\fi
\ifx \bauthor  \undefined \def \bauthor#1{#1}\fi
\ifx \batitle  \undefined \def \batitle#1{#1}\fi
\ifx \bjtitle  \undefined \def \bjtitle#1{#1}\fi
\ifx \bvolume  \undefined \def \bvolume#1{\textbf{#1}}\fi
\ifx \byear  \undefined \def \byear#1{#1}\fi
\ifx \bissue  \undefined \def \bissue#1{#1}\fi
\ifx \bfpage  \undefined \def \bfpage#1{#1}\fi
\ifx \blpage  \undefined \def \blpage #1{#1}\fi
\ifx \burl  \undefined \def \burl#1{\textsf{#1}}\fi
\ifx \doiurl  \undefined \def \doiurl#1{\url{https://doi.org/#1}}\fi
\ifx \betal  \undefined \def \betal{\textit{et al.}}\fi
\ifx \binstitute  \undefined \def \binstitute#1{#1}\fi
\ifx \binstitutionaled  \undefined \def \binstitutionaled#1{#1}\fi
\ifx \bctitle  \undefined \def \bctitle#1{#1}\fi
\ifx \beditor  \undefined \def \beditor#1{#1}\fi
\ifx \bpublisher  \undefined \def \bpublisher#1{#1}\fi
\ifx \bbtitle  \undefined \def \bbtitle#1{#1}\fi
\ifx \bedition  \undefined \def \bedition#1{#1}\fi
\ifx \bseriesno  \undefined \def \bseriesno#1{#1}\fi
\ifx \blocation  \undefined \def \blocation#1{#1}\fi
\ifx \bsertitle  \undefined \def \bsertitle#1{#1}\fi
\ifx \bsnm \undefined \def \bsnm#1{#1}\fi
\ifx \bsuffix \undefined \def \bsuffix#1{#1}\fi
\ifx \bparticle \undefined \def \bparticle#1{#1}\fi
\ifx \barticle \undefined \def \barticle#1{#1}\fi
\bibcommenthead
\ifx \bconfdate \undefined \def \bconfdate #1{#1}\fi
\ifx \botherref \undefined \def \botherref #1{#1}\fi
\ifx \url \undefined \def \url#1{\textsf{#1}}\fi
\ifx \bchapter \undefined \def \bchapter#1{#1}\fi
\ifx \bbook \undefined \def \bbook#1{#1}\fi
\ifx \bcomment \undefined \def \bcomment#1{#1}\fi
\ifx \oauthor \undefined \def \oauthor#1{#1}\fi
\ifx \citeauthoryear \undefined \def \citeauthoryear#1{#1}\fi
\ifx \endbibitem  \undefined \def \endbibitem {}\fi
\ifx \bconflocation  \undefined \def \bconflocation#1{#1}\fi
\ifx \arxivurl  \undefined \def \arxivurl#1{\textsf{#1}}\fi
\csname PreBibitemsHook\endcsname

\bibitem[\protect\citeauthoryear{Lalo{\"e}}{2019}]{laloë2019}
\begin{bbook}
\bauthor{\bsnm{Lalo{\"e}}, \binits{F.}}:
\bbtitle{Do We Really Understand Quantum Mechanics?},
pp. \bfpage{189}--\blpage{222}.
\bpublisher{Cambridge University Press},
\blocation{Cambridge}
(\byear{2019}).
\doiurl{10.1017/9781108569361.010}
\end{bbook}
\endbibitem

\bibitem[\protect\citeauthoryear{Blasiak and Markiewicz}{2019}]{blasiak2019}
\begin{barticle}
\bauthor{\bsnm{Blasiak}, \binits{P.}},
\bauthor{\bsnm{Markiewicz}, \binits{M.}}:
\batitle{Entangling three qubits without ever touching}.
\bjtitle{Scientific Reports}
\bvolume{9}(\bissue{1}),
\bfpage{20131}
(\byear{2019})
\end{barticle}
\endbibitem

\bibitem[\protect\citeauthoryear{Das et~al.}{2022}]{das2022}
\begin{barticle}
\bauthor{\bsnm{Das}, \binits{T.}},
\bauthor{\bsnm{Karczewski}, \binits{M.}},
\bauthor{\bsnm{Mandarino}, \binits{A.}},
\bauthor{\bsnm{Markiewicz}, \binits{M.}},
\bauthor{\bsnm{Woloncewicz}, \binits{B.}},
\bauthor{\bsnm{{\.Z}ukowski}, \binits{M.}}:
\batitle{Comment on ‘single particle nonlocality with completely independent reference states’}.
\bjtitle{New Journal of Physics}
\bvolume{24}(\bissue{3}),
\bfpage{038001}
(\byear{2022})
\end{barticle}
\endbibitem

\bibitem[\protect\citeauthoryear{Mooney et~al.}{2019}]{mooney2019}
\begin{barticle}
\bauthor{\bsnm{Mooney}, \binits{G.J.}},
\bauthor{\bsnm{Hill}, \binits{C.D.}},
\bauthor{\bsnm{Hollenberg}, \binits{L.C.}}:
\batitle{Entanglement in a 20-qubit superconducting quantum computer}.
\bjtitle{Scientific reports}
\bvolume{9}(\bissue{1}),
\bfpage{13465}
(\byear{2019})
\end{barticle}
\endbibitem

\bibitem[\protect\citeauthoryear{Srinivas et~al.}{2021}]{srinivas2021highfidelity}
\begin{barticle}
\bauthor{\bsnm{Srinivas}, \binits{R.}},
\bauthor{\bsnm{Knill}, \binits{E.}},
\bauthor{\bsnm{Sutherland}, \binits{R.}},
\bauthor{\bsnm{Kwiatkowski}, \binits{A.T.}},
\bauthor{\bsnm{Knaack}, \binits{H.M.}},
\bauthor{\bsnm{Glancy}, \binits{S.}},
\bauthor{\bsnm{Wineland}, \binits{D.J.}},
\bauthor{\bsnm{Burd}, \binits{S.C.}},
\bauthor{\bsnm{Leibfried}, \binits{D.}},
\bauthor{\bsnm{Wilson}, \binits{A.C.}},
\bauthor{\bsnm{Allcock}, \binits{D.T.}},
\bauthor{\bsnm{Slichter}, \binits{D.}}:
\batitle{High-fidelity laser-free universal control of trapped ion qubits}.
\bjtitle{Nature}
\bvolume{597},
\bfpage{209}--\blpage{213}
(\byear{2021})
\doiurl{10.1038/s41586-021-03809-4}
\end{barticle}
\endbibitem

\bibitem[\protect\citeauthoryear{Yin et~al.}{2020}]{Yin2020Entangl}
\begin{barticle}
\bauthor{\bsnm{Yin}, \binits{J.}},
\bauthor{\bsnm{Li}, \binits{Y.}},
\bauthor{\bsnm{Liao}, \binits{S.}},
\bauthor{\bsnm{Yang}, \binits{M.}},
\bauthor{\bsnm{Cao}, \binits{Y.}},
\bauthor{\bsnm{Zhang}, \binits{L.}},
\bauthor{\bsnm{Ren}, \binits{J.-G.}},
\bauthor{\bsnm{Cai}, \binits{W.}},
\bauthor{\bsnm{Liu}, \binits{W.}},
\bauthor{\bsnm{Li}, \binits{S.-L.}},
\bauthor{\bsnm{Shu}, \binits{R.}},
\bauthor{\bsnm{Huang}, \binits{Y.}},
\bauthor{\bsnm{Deng}, \binits{L.}},
\bauthor{\bsnm{Li}, \binits{L.}},
\bauthor{\bsnm{Zhang}, \binits{Q.}},
\bauthor{\bsnm{Liu}, \binits{N.-L.}},
\bauthor{\bsnm{Chen}, \binits{Y.-A.}},
\bauthor{\bsnm{Lu}, \binits{C.}},
\bauthor{\bsnm{Wang}, \binits{X.-B.}},
\bauthor{\bsnm{Xu}, \binits{F.}},
\bauthor{\bsnm{Wang}, \binits{J.-Y.}},
\bauthor{\bsnm{Peng}, \binits{C.-Z.}},
\bauthor{\bsnm{Ekert}, \binits{A.}},
\bauthor{\bsnm{Pan}, \binits{J.-W.}}:
\batitle{Entanglement-based secure quantum cryptography over 1,120 kilometres}.
\bjtitle{Nature}
\bvolume{582},
\bfpage{501}--\blpage{505}
(\byear{2020})
\doiurl{10.1038/s41586-020-2401-y}
\end{barticle}
\endbibitem

\bibitem[\protect\citeauthoryear{Morstyn}{2022}]{morstyn2022annealing}
\begin{barticle}
\bauthor{\bsnm{Morstyn}, \binits{T.}}:
\batitle{Annealing-based quantum computing for combinatorial optimal power flow}.
\bjtitle{IEEE Transactions on Smart Grid}
\bvolume{14}(\bissue{2}),
\bfpage{1093}--\blpage{1102}
(\byear{2022})
\end{barticle}
\endbibitem

\bibitem[\protect\citeauthoryear{Díez-Valle et~al.}{2021}]{diezvalle2021quantum}
\begin{barticle}
\bauthor{\bsnm{Díez-Valle}, \binits{P.}},
\bauthor{\bsnm{Porras}, \binits{D.}},
\bauthor{\bsnm{García-Ripoll}, \binits{J.J.}}:
\batitle{Quantum variational optimization: The role of entanglement and problem hardness}.
\bjtitle{Physical Review A}
\bvolume{104}(\bissue{6}),
\bfpage{062426}
(\byear{2021})
\doiurl{10.1103/PhysRevA.104.062426}
\end{barticle}
\endbibitem

\bibitem[\protect\citeauthoryear{Liu et~al.}{2022}]{Liu2022Detecting}
\begin{barticle}
\bauthor{\bsnm{Liu}, \binits{Z.}},
\bauthor{\bsnm{Tang}, \binits{Y.}},
\bauthor{\bsnm{Dai}, \binits{H.}},
\bauthor{\bsnm{Liu}, \binits{P.}},
\bauthor{\bsnm{Chen}, \binits{S.}},
\bauthor{\bsnm{Ma}, \binits{X.}}:
\batitle{Detecting entanglement in quantum many-body systems via permutation moments}.
\bjtitle{Phys. Rev. Lett.}
\bvolume{129},
\bfpage{260501}
(\byear{2022})
\doiurl{10.1103/PhysRevLett.129.260501}
\end{barticle}
\endbibitem

\bibitem[\protect\citeauthoryear{Fadel et~al.}{2020}]{PhysRevA.101.052117}
\begin{barticle}
\bauthor{\bsnm{Fadel}, \binits{M.}},
\bauthor{\bsnm{Ares}, \binits{L.}},
\bauthor{\bsnm{Luis}, \binits{A.}},
\bauthor{\bsnm{He}, \binits{Q.}}:
\batitle{Number-phase entanglement and einstein-podolsky-rosen steering}.
\bjtitle{Phys. Rev. A}
\bvolume{101},
\bfpage{052117}
(\byear{2020})
\doiurl{10.1103/PhysRevA.101.052117}
\end{barticle}
\endbibitem

\bibitem[\protect\citeauthoryear{Zangi et~al.}{2021}]{Zangi2021Combo}
\begin{botherref}
\oauthor{\bsnm{Zangi}, \binits{S.}},
\oauthor{\bsnm{Wu}, \binits{J.}},
\oauthor{\bsnm{Qiao}, \binits{C.}}:
Combo separability criteria and lower bound on concurrence.
Journal of Physics A: Mathematical and Theoretical
\textbf{55}
(2021)
\doiurl{10.1088/1751-8121/ac3c80}
\end{botherref}
\endbibitem

\bibitem[\protect\citeauthoryear{Huber et~al.}{2018}]{Huber2018High}
\begin{barticle}
\bauthor{\bsnm{Huber}, \binits{M.}},
\bauthor{\bsnm{Lami}, \binits{L.}},
\bauthor{\bsnm{Lancien}, \binits{C.}},
\bauthor{\bsnm{Müller-Hermes}, \binits{A.}}:
\batitle{High-dimensional entanglement in states with positive partial transposition.}
\bjtitle{Physical review letters}
\bvolume{121 20},
\bfpage{200503}
(\byear{2018})
\doiurl{10.1103/PhysRevLett.121.200503}
\end{barticle}
\endbibitem

\bibitem[\protect\citeauthoryear{Bai et~al.}{2014}]{Bai2014General}
\begin{barticle}
\bauthor{\bsnm{Bai}, \binits{Y.-K.}},
\bauthor{\bsnm{Xu}, \binits{Y.-F.}},
\bauthor{\bsnm{Wang}, \binits{Z.D.}}:
\batitle{General monogamy relation for the entanglement of formation in multiqubit systems.}
\bjtitle{Physical review letters}
\bvolume{113 10},
\bfpage{100503}
(\byear{2014})
\doiurl{10.1103/PhysRevLett.113.100503}
\end{barticle}
\endbibitem

\bibitem[\protect\citeauthoryear{Kim}{2021}]{Kim2021Entanglement}
\begin{botherref}
\oauthor{\bsnm{Kim}, \binits{J.S.}}:
Entanglement of formation and monogamy of multi-party quantum entanglement.
Scientific Reports
\textbf{11}
(2021)
\doiurl{10.1038/s41598-021-82052-3}
\end{botherref}
\endbibitem

\bibitem[\protect\citeauthoryear{Wang et~al.}{2022}]{Wang2022}
\begin{barticle}
\bauthor{\bsnm{Wang}, \binits{K.}},
\bauthor{\bsnm{Song}, \binits{Z.}},
\bauthor{\bsnm{Zhao}, \binits{X.}},
\bauthor{\bsnm{Wang}, \binits{Z.}},
\bauthor{\bsnm{Wang}, \binits{X.}}:
\batitle{Detecting and quantifying entanglement on near-term quantum devices}.
\bjtitle{npj Quantum Information}
(\byear{2022})
\doiurl{10.1038/s41534-022-00556-w}
\end{barticle}
\endbibitem

\bibitem[\protect\citeauthoryear{Bhaskara and Panigrahi}{2016}]{Bhaskara2016Generalized}
\begin{barticle}
\bauthor{\bsnm{Bhaskara}, \binits{V.S.}},
\bauthor{\bsnm{Panigrahi}, \binits{P.}}:
\batitle{Generalized concurrence measure for faithful quantification of multiparticle pure state entanglement using lagrange’s identity and wedge product}.
\bjtitle{Quantum Information Processing}
\bvolume{16},
\bfpage{1}--\blpage{15}
(\byear{2016})
\doiurl{10.1007/s11128-017-1568-0}
\end{barticle}
\endbibitem

\bibitem[\protect\citeauthoryear{Arkhipov et~al.}{2018}]{arkhipov2018}
\begin{barticle}
\bauthor{\bsnm{Arkhipov}, \binits{I.I.}},
\bauthor{\bsnm{Barasi{\'n}ski}, \binits{A.}},
\bauthor{\bsnm{Svozil{\'\i}k}, \binits{J.}}:
\batitle{Negativity volume of the generalized wigner function as an entanglement witness for hybrid bipartite states}.
\bjtitle{Scientific reports}
\bvolume{8}(\bissue{1}),
\bfpage{16955}
(\byear{2018})
\end{barticle}
\endbibitem

\bibitem[\protect\citeauthoryear{Ma and Yung}{2018}]{ma2018}
\begin{barticle}
\bauthor{\bsnm{Ma}, \binits{Y.-C.}},
\bauthor{\bsnm{Yung}, \binits{M.-H.}}:
\batitle{Transforming bell’s inequalities into state classifiers with machine learning}.
\bjtitle{npj Quantum Information}
\bvolume{4}(\bissue{1}),
\bfpage{34}
(\byear{2018})
\end{barticle}
\endbibitem

\bibitem[\protect\citeauthoryear{Lu et~al.}{2018}]{lu2018}
\begin{barticle}
\bauthor{\bsnm{Lu}, \binits{S.}},
\bauthor{\bsnm{Huang}, \binits{S.}},
\bauthor{\bsnm{Li}, \binits{K.}},
\bauthor{\bsnm{Li}, \binits{J.}},
\bauthor{\bsnm{Chen}, \binits{J.}},
\bauthor{\bsnm{Lu}, \binits{D.}},
\bauthor{\bsnm{Ji}, \binits{Z.}},
\bauthor{\bsnm{Shen}, \binits{Y.}},
\bauthor{\bsnm{Zhou}, \binits{D.}},
\bauthor{\bsnm{Zeng}, \binits{B.}}:
\batitle{Separability-entanglement classifier via machine learning}.
\bjtitle{Physical Review A}
\bvolume{98}(\bissue{1}),
\bfpage{012315}
(\byear{2018})
\end{barticle}
\endbibitem

\bibitem[\protect\citeauthoryear{Hyllus and Eisert}{2006}]{hyllus2006}
\begin{barticle}
\bauthor{\bsnm{Hyllus}, \binits{P.}},
\bauthor{\bsnm{Eisert}, \binits{J.}}:
\batitle{Optimal entanglement witnesses for continuous-variable systems}.
\bjtitle{New Journal of Physics}
\bvolume{8}(\bissue{4}),
\bfpage{51}
(\byear{2006})
\end{barticle}
\endbibitem

\bibitem[\protect\citeauthoryear{Qi and Hou}{2012}]{qi2012}
\begin{barticle}
\bauthor{\bsnm{Qi}, \binits{X.}},
\bauthor{\bsnm{Hou}, \binits{J.}}:
\batitle{Characterization of optimal entanglement witnesses}.
\bjtitle{Physical Review A}
\bvolume{85}(\bissue{2}),
\bfpage{022334}
(\byear{2012})
\end{barticle}
\endbibitem

\bibitem[\protect\citeauthoryear{Li et~al.}{2020}]{Li2020A}
\begin{barticle}
\bauthor{\bsnm{Li}, \binits{Z.}},
\bauthor{\bsnm{Liu}, \binits{F.}},
\bauthor{\bsnm{Yang}, \binits{W.}},
\bauthor{\bsnm{Peng}, \binits{S.}},
\bauthor{\bsnm{Zhou}, \binits{J.}}:
\batitle{A survey of convolutional neural networks: Analysis, applications, and prospects}.
\bjtitle{IEEE Transactions on Neural Networks and Learning Systems}
\bvolume{33},
\bfpage{6999}--\blpage{7019}
(\byear{2020})
\doiurl{10.1109/TNNLS.2021.3084827}
\end{barticle}
\endbibitem

\bibitem[\protect\citeauthoryear{Ahmed et~al.}{2020}]{Ahmed2020ClassificationAR}
\begin{botherref}
\oauthor{\bsnm{Ahmed}, \binits{S.}},
\oauthor{\bsnm{Mu{\~n}oz}, \binits{C.S.}},
\oauthor{\bsnm{Nori}, \binits{F.}},
\oauthor{\bsnm{Kockum}, \binits{A.F.}}:
Classification and reconstruction of optical quantum states with deep neural networks.
ArXiv
\textbf{abs/2012.02185}
(2020)
\end{botherref}
\endbibitem

\bibitem[\protect\citeauthoryear{Qiu et~al.}{2019}]{qiu2019}
\begin{barticle}
\bauthor{\bsnm{Qiu}, \binits{P.-H.}},
\bauthor{\bsnm{Chen}, \binits{X.-G.}},
\bauthor{\bsnm{Shi}, \binits{Y.-W.}}:
\batitle{Detecting entanglement with deep quantum neural networks}.
\bjtitle{IEEE Access}
\bvolume{7},
\bfpage{94310}--\blpage{94320}
(\byear{2019})
\end{barticle}
\endbibitem

\bibitem[\protect\citeauthoryear{Harney et~al.}{2021}]{harney2021}
\begin{barticle}
\bauthor{\bsnm{Harney}, \binits{C.}},
\bauthor{\bsnm{Paternostro}, \binits{M.}},
\bauthor{\bsnm{Pirandola}, \binits{S.}}:
\batitle{Mixed state entanglement classification using artificial neural networks}.
\bjtitle{New Journal of Physics}
\bvolume{23}(\bissue{6}),
\bfpage{063033}
(\byear{2021})
\end{barticle}
\endbibitem

\bibitem[\protect\citeauthoryear{Girardin et~al.}{2022}]{girardin2022}
\begin{barticle}
\bauthor{\bsnm{Girardin}, \binits{A.}},
\bauthor{\bsnm{Brunner}, \binits{N.}},
\bauthor{\bsnm{Kriv{\'a}chy}, \binits{T.}}:
\batitle{Building separable approximations for quantum states via neural networks}.
\bjtitle{Physical Review Research}
\bvolume{4}(\bissue{2}),
\bfpage{023238}
(\byear{2022})
\end{barticle}
\endbibitem

\bibitem[\protect\citeauthoryear{Chalumuri et~al.}{2021}]{Chalumuri2021}
\begin{botherref}
\oauthor{\bsnm{Chalumuri}, \binits{A.}},
\oauthor{\bsnm{Kune}, \binits{R.}},
\oauthor{\bsnm{Manoj}, \binits{B.S.}}:
A hybrid classical-quantum approach for multi-class classification.
Quantum Information Processing
\textbf{20}(3)
(2021)
\doiurl{10.1007/s11128-021-03029-9}
\end{botherref}
\endbibitem

\bibitem[\protect\citeauthoryear{Fanizza et~al.}{2022}]{Fanizza2022}
\begin{barticle}
\bauthor{\bsnm{Fanizza}, \binits{M.}},
\bauthor{\bsnm{Skotiniotis}, \binits{M.}},
\bauthor{\bsnm{Calsamiglia}, \binits{J.}},
\bauthor{\bsnm{Muñoz-Tapia}, \binits{R.}},
\bauthor{\bsnm{Sentís}, \binits{G.}}:
\batitle{Universal algorithms for quantum data learning}.
\bjtitle{Europhysics Letters}
(\byear{2022})
\doiurl{10.1209/0295-5075/ac9c29}
\end{barticle}
\endbibitem

\bibitem[\protect\citeauthoryear{Paini et~al.}{2021}]{paini2021}
\begin{barticle}
\bauthor{\bsnm{Paini}, \binits{M.}},
\bauthor{\bsnm{Kalev}, \binits{A.}},
\bauthor{\bsnm{Padilha}, \binits{D.}},
\bauthor{\bsnm{Ruck}, \binits{B.}}:
\batitle{Estimating expectation values using approximate quantum states}.
\bjtitle{Quantum}
\bvolume{5},
\bfpage{413}
(\byear{2021})
\end{barticle}
\endbibitem

\bibitem[\protect\citeauthoryear{Gu et~al.}{2019}]{gu2019}
\begin{barticle}
\bauthor{\bsnm{Gu}, \binits{X.}},
\bauthor{\bsnm{Chen}, \binits{L.}},
\bauthor{\bsnm{Zeilinger}, \binits{A.}},
\bauthor{\bsnm{Krenn}, \binits{M.}}:
\batitle{Quantum experiments and graphs. iii. high-dimensional and multiparticle entanglement}.
\bjtitle{Physical Review A}
\bvolume{99}(\bissue{3}),
\bfpage{032338}
(\byear{2019})
\end{barticle}
\endbibitem

\bibitem[\protect\citeauthoryear{Johansson et~al.}{2013}]{qutip2}
\begin{barticle}
\bauthor{\bsnm{Johansson}, \binits{J.R.}},
\bauthor{\bsnm{Nation}, \binits{P.D.}},
\bauthor{\bsnm{Nori}, \binits{F.}}:
\batitle{Qutip 2: A python framework for the dynamics of open quantum systems}.
\bjtitle{Computer Physics Communications}
\bvolume{184},
\bfpage{1234}
(\byear{2013})
\doiurl{10.1016/j.cpc.2012.11.019}
\end{barticle}
\endbibitem

\bibitem[\protect\citeauthoryear{Kookani}{2022}]{alikookani_2022}
\begin{botherref}
\oauthor{\bsnm{Kookani}, \binits{A.}}:
Quantangle.
Kaggle
(2022).
\doiurl{10.34740/KAGGLE/DSV/3663663} .
\url{https://www.kaggle.com/dsv/3663663}
\end{botherref}
\endbibitem

\bibitem[\protect\citeauthoryear{Team}{2020}]{Qiskit}
\begin{botherref}
\oauthor{\bsnm{Team}, \binits{Q.D.}}:
Qiskit: An Open-source Framework for Quantum Computing.
\url{https://qiskit.org}
(2020).
\doiurl{10.5281/zenodo.2562110}
\end{botherref}
\endbibitem

\bibitem[\protect\citeauthoryear{Chollet}{2021}]{chollet2021learning}
\begin{bbook}
\bauthor{\bsnm{Chollet}, \binits{F.}}:
\bbtitle{Deep Learning with Python}.
\bpublisher{Simon and Schuster},
\blocation{New York}
(\byear{2021})
\end{bbook}
\endbibitem

\bibitem[\protect\citeauthoryear{Jia et~al.}{2022}]{jia2022}
\begin{barticle}
\bauthor{\bsnm{Jia}, \binits{L.}},
\bauthor{\bsnm{Ga{\"u}z{\`e}re}, \binits{B.}},
\bauthor{\bsnm{Honeine}, \binits{P.}}:
\batitle{Graph kernels based on linear patterns: theoretical and experimental comparisons}.
\bjtitle{Expert Systems with Applications}
\bvolume{189},
\bfpage{116095}
(\byear{2022})
\end{barticle}
\endbibitem

\bibitem[\protect\citeauthoryear{Santurkar et~al.}{2018}]{santurkar2018}
\begin{botherref}
\oauthor{\bsnm{Santurkar}, \binits{S.}},
\oauthor{\bsnm{Tsipras}, \binits{D.}},
\oauthor{\bsnm{Ilyas}, \binits{A.}},
\oauthor{\bsnm{Madry}, \binits{A.}}:
How does batch normalization help optimization?
Advances in neural information processing systems
\textbf{31}
(2018)
\end{botherref}
\endbibitem

\bibitem[\protect\citeauthoryear{Liu et~al.}{2020}]{liu2020}
\begin{bchapter}
\bauthor{\bsnm{Liu}, \binits{J.-J.}},
\bauthor{\bsnm{Hou}, \binits{Q.}},
\bauthor{\bsnm{Cheng}, \binits{M.-M.}},
\bauthor{\bsnm{Wang}, \binits{C.}},
\bauthor{\bsnm{Feng}, \binits{J.}}:
\bctitle{Improving convolutional networks with self-calibrated convolutions}.
In: \bbtitle{Proceedings of the IEEE/CVF Conference on Computer Vision and Pattern Recognition},
pp. \bfpage{10096}--\blpage{10105}
(\byear{2020})
\end{bchapter}
\endbibitem

\bibitem[\protect\citeauthoryear{Yoshida and Okada}{2020}]{Yoshida2020Data-dependence}
\begin{botherref}
\oauthor{\bsnm{Yoshida}, \binits{Y.}},
\oauthor{\bsnm{Okada}, \binits{M.}}:
Data-dependence of plateau phenomenon in learning with neural network—statistical mechanical analysis.
Journal of Statistical Mechanics: Theory and Experiment
\textbf{2020}
(2020)
\doiurl{10.1088/1742-5468/abc62f}
\end{botherref}
\endbibitem

\bibitem[\protect\citeauthoryear{Lan et~al.}{2019}]{Lan2019Image}
\begin{barticle}
\bauthor{\bsnm{Lan}, \binits{R.}},
\bauthor{\bsnm{Zou}, \binits{H.}},
\bauthor{\bsnm{Pang}, \binits{C.}},
\bauthor{\bsnm{Zhong}, \binits{Y.}},
\bauthor{\bsnm{Liu}, \binits{Z.}},
\bauthor{\bsnm{Luo}, \binits{X.}}:
\batitle{Image denoising via deep residual convolutional neural networks}.
\bjtitle{Signal, Image and Video Processing}
\bvolume{15},
\bfpage{1}--\blpage{8}
(\byear{2019})
\doiurl{10.1007/S11760-019-01537-X}
\end{barticle}
\endbibitem

\bibitem[\protect\citeauthoryear{Givi et~al.}{2015}]{Givi2015Modelling}
\begin{barticle}
\bauthor{\bsnm{Givi}, \binits{Z.S.}},
\bauthor{\bsnm{Jaber}, \binits{M.}},
\bauthor{\bsnm{Neumann}, \binits{W.}}:
\batitle{Modelling worker reliability with learning and fatigue}.
\bjtitle{Applied Mathematical Modelling}
\bvolume{39},
\bfpage{5186}--\blpage{5199}
(\byear{2015})
\doiurl{10.1016/J.APM.2015.03.038}
\end{barticle}
\endbibitem

\bibitem[\protect\citeauthoryear{Yu et~al.}{2020}]{Yu2020LLR}
\begin{barticle}
\bauthor{\bsnm{Yu}, \binits{C.}},
\bauthor{\bsnm{Qi}, \binits{X.}},
\bauthor{\bsnm{Ma}, \binits{H.}},
\bauthor{\bsnm{He}, \binits{X.}},
\bauthor{\bsnm{Wang}, \binits{C.}},
\bauthor{\bsnm{Zhao}, \binits{Y.}}:
\batitle{Llr: Learning learning rates by lstm for training neural networks}.
\bjtitle{Neurocomputing}
\bvolume{394},
\bfpage{41}--\blpage{50}
(\byear{2020})
\doiurl{10.1016/j.neucom.2020.01.106}
\end{barticle}
\endbibitem

\bibitem[\protect\citeauthoryear{Virzì et~al.}{2019}]{Virz2019}
\begin{barticle}
\bauthor{\bsnm{Virzì}, \binits{S.}},
\bauthor{\bsnm{Rebufello}, \binits{E.}},
\bauthor{\bsnm{Avella}, \binits{A.}},
\bauthor{\bsnm{Piacentini}, \binits{F.}},
\bauthor{\bsnm{Gramegna}, \binits{M.}},
\bauthor{\bsnm{Berchera}, \binits{I.R.}},
\bauthor{\bsnm{Degiovanni}, \binits{I.P.}},
\bauthor{\bsnm{Genovese}, \binits{M.}}:
\batitle{Optimal estimation of entanglement and discord in two-qubit states}.
\bjtitle{Scientific Reports}
\bvolume{9},
\bfpage{3030}
(\byear{2019})
\doiurl{10.1038/s41598-019-39334-8}
\end{barticle}
\endbibitem

\bibitem[\protect\citeauthoryear{Lanyon et~al.}{2016}]{Lanyon2016Efficient}
\begin{barticle}
\bauthor{\bsnm{Lanyon}, \binits{B.}},
\bauthor{\bsnm{Maier}, \binits{C.}},
\bauthor{\bsnm{Holzapfel}, \binits{M.}},
\bauthor{\bsnm{Baumgratz}, \binits{T.}},
\bauthor{\bsnm{Hempel}, \binits{C.}},
\bauthor{\bsnm{Jurcevic}, \binits{P.}},
\bauthor{\bsnm{Dhand}, \binits{I.}},
\bauthor{\bsnm{Buyskikh}, \binits{A.}},
\bauthor{\bsnm{Daley}, \binits{A.}},
\bauthor{\bsnm{Cramer}, \binits{M.}},
\bauthor{\bsnm{Plenio}, \binits{M.}},
\bauthor{\bsnm{Blatt}, \binits{R.}},
\bauthor{\bsnm{Roos}, \binits{C.}}:
\batitle{Efficient tomography of a quantum many-body system}.
\bjtitle{Nature Physics}
\bvolume{13},
\bfpage{1158}--\blpage{1162}
(\byear{2016})
\doiurl{10.1038/nphys4244}
\end{barticle}
\endbibitem

\bibitem[\protect\citeauthoryear{Nielsen and Chuang}{2001}]{nielsen2001quantum}
\begin{bbook}
\bauthor{\bsnm{Nielsen}, \binits{M.A.}},
\bauthor{\bsnm{Chuang}, \binits{I.L.}}:
\bbtitle{Quantum Computation and Quantum Information}
vol. \bseriesno{2},
pp. \bfpage{389}--\blpage{393}.
\bpublisher{Cambridge university press Cambridge},
\blocation{Cambridge}
(\byear{2001})
\end{bbook}
\endbibitem

\bibitem[\protect\citeauthoryear{Essakhi et~al.}{2022}]{essakhi2022intrinsic}
\begin{barticle}
\bauthor{\bsnm{Essakhi}, \binits{M.}},
\bauthor{\bsnm{Khedif}, \binits{Y.}},
\bauthor{\bsnm{Mansour}, \binits{M.}},
\bauthor{\bsnm{Daoud}, \binits{M.}}:
\batitle{Intrinsic decoherence effects on quantum correlations dynamics}.
\bjtitle{Optical and Quantum Electronics}
\bvolume{54},
\bfpage{1}--\blpage{15}
(\byear{2022})
\end{barticle}
\endbibitem

\bibitem[\protect\citeauthoryear{Facchi et~al.}{2019}]{Facchi2019Phase}
\begin{botherref}
\oauthor{\bsnm{Facchi}, \binits{P.}},
\oauthor{\bsnm{Parisi}, \binits{G.}},
\oauthor{\bsnm{Pascazio}, \binits{S.}},
\oauthor{\bsnm{Scardicchio}, \binits{A.}},
\oauthor{\bsnm{Yuasa}, \binits{K.}}:
Phase diagram of bipartite entanglement.
Journal of Physics A: Mathematical and Theoretical
\textbf{52}
(2019)
\doiurl{10.1088/1751-8121/ab3f4e}
\end{botherref}
\endbibitem

\bibitem[\protect\citeauthoryear{Yosefpor et~al.}{2020}]{Yosefpor_2020}
\begin{barticle}
\bauthor{\bsnm{Yosefpor}, \binits{M.}},
\bauthor{\bsnm{Mostaan}, \binits{M.R.}},
\bauthor{\bsnm{Raeisi}, \binits{S.}}:
\batitle{Finding semi-optimal measurements for entanglement detection using autoencoder neural networks}.
\bjtitle{Quantum Science and Technology}
\bvolume{5}(\bissue{4}),
\bfpage{045006}
(\byear{2020})
\doiurl{10.1088/2058-9565/aba34c}
\end{barticle}
\endbibitem

\bibitem[\protect\citeauthoryear{Chen et~al.}{2021}]{chen2021}
\begin{barticle}
\bauthor{\bsnm{Chen}, \binits{Y.}},
\bauthor{\bsnm{Pan}, \binits{Y.}},
\bauthor{\bsnm{Zhang}, \binits{G.}},
\bauthor{\bsnm{Cheng}, \binits{S.}}:
\batitle{Detecting quantum entanglement with unsupervised learning}.
\bjtitle{Quantum Science and Technology}
\bvolume{7}(\bissue{1}),
\bfpage{015005}
(\byear{2021})
\end{barticle}
\endbibitem

\bibitem[\protect\citeauthoryear{Asif et~al.}{2023}]{asif2023}
\begin{barticle}
\bauthor{\bsnm{Asif}, \binits{N.}},
\bauthor{\bsnm{Khalid}, \binits{U.}},
\bauthor{\bsnm{Khan}, \binits{A.}},
\bauthor{\bsnm{Duong}, \binits{T.Q.}},
\bauthor{\bsnm{Shin}, \binits{H.}}:
\batitle{Entanglement detection with artificial neural networks}.
\bjtitle{Scientific Reports}
\bvolume{13}(\bissue{1}),
\bfpage{1562}
(\byear{2023})
\end{barticle}
\endbibitem

\bibitem[\protect\citeauthoryear{Qu et~al.}{2023}]{Qu2023EntanglementDW}
\begin{barticle}
\bauthor{\bsnm{Qu}, \binits{Y.-D.}},
\bauthor{\bsnm{Zhang}, \binits{R.-Q.}},
\bauthor{\bsnm{Shen}, \binits{S.}},
\bauthor{\bsnm{Yu}, \binits{J.}},
\bauthor{\bsnm{Li}, \binits{M.}}:
\batitle{Entanglement detection with complex-valued neural networks}.
\bjtitle{International Journal of Theoretical Physics}
\bvolume{62},
\bfpage{1}--\blpage{15}
(\byear{2023})
\end{barticle}
\endbibitem

\bibitem[\protect\citeauthoryear{Paw{\l}owski and Krawczyk}{2022}]{pawlowski2022quantification}
\begin{botherref}
\oauthor{\bsnm{Paw{\l}owski}, \binits{J.}},
\oauthor{\bsnm{Krawczyk}, \binits{M.}}:
Quantification of entanglement with siamese convolutional neural networks.
arXiv preprint arXiv:2210.07410
(2022)
\end{botherref}
\endbibitem

\bibitem[\protect\citeauthoryear{Xie and Eberly}{2021}]{xie2021triangle}
\begin{barticle}
\bauthor{\bsnm{Xie}, \binits{S.}},
\bauthor{\bsnm{Eberly}, \binits{J.H.}}:
\batitle{Triangle measure of tripartite entanglement}.
\bjtitle{Phys. Rev. Lett.}
\bvolume{127},
\bfpage{040403}
(\byear{2021})
\doiurl{10.1103/PhysRevLett.127.040403}
\end{barticle}
\endbibitem

\bibitem[\protect\citeauthoryear{Baccari et~al.}{2017}]{baccari2017efficient}
\begin{barticle}
\bauthor{\bsnm{Baccari}, \binits{F.}},
\bauthor{\bsnm{Cavalcanti}, \binits{D.}},
\bauthor{\bsnm{Wittek}, \binits{P.}},
\bauthor{\bsnm{Acín}, \binits{A.}}:
\batitle{Efficient device-independent entanglement detection for multipartite systems}.
\bjtitle{Phys. Rev. X}
\bvolume{7},
\bfpage{021042}
(\byear{2017})
\doiurl{10.1103/PhysRevX.7.021042}
\end{barticle}
\endbibitem

\end{thebibliography}

\end{document}